\documentclass[twocolumn, floatfix]{aastex631}
\usepackage{amsmath} % or simply amstext

\usepackage{comment}
\usepackage{outlines}
\usepackage{color}

\usepackage{appendix}
\usepackage{multirow}

\begin{document}

\title{Cross-correlation of Luminous Red Galaxies with ML-selected AGN in HSC-SSP II: \\ AGN classification and clustering with DESI spectroscopy}
\shorttitle{AGN classification and clustering with DESI spectroscopy}
\shortauthors{C\'ordova~Rosado et al.}

%% Note that the corresponding author command and emails has to come
%% before everything else. Also place all the emails in the \email
%% command instead of using multiple \email calls.
\author[0000-0002-7967-7676]{Rodrigo~C\'ordova~Rosado}
\affiliation{Department of Astrophysical Sciences, Peyton Hall, Princeton University, 4 Ivy Lane, Princeton, NJ 08544, USA}
\author[0000-0003-4700-663X]{Andy~D.~Goulding}
\affiliation{Department of Astrophysical Sciences, Peyton Hall, Princeton University, 4 Ivy Lane, Princeton, NJ 08544, USA}
\author[0000-0002-5612-3427]{Jenny~E.~Greene}
\affiliation{Department of Astrophysical Sciences, Peyton Hall, Princeton University, 4 Ivy Lane, Princeton, NJ 08544, USA}

\author[0000-0002-5808-4708]{Nickolas~Kokron}
\affiliation{Department of Astrophysical Sciences, Peyton Hall, Princeton University, 4 Ivy Lane, Princeton, NJ 08544, USA}

\affiliation{School of Natural Sciences, Institute for Advanced Study, 1 Einstein Drive, Princeton, NJ, 08540, USA}

\author[0000-0002-0106-7755]{Michael~A.~Strauss}
\affiliation{Department of Astrophysical Sciences, Peyton Hall, Princeton University, 4 Ivy Lane, Princeton, NJ 08544, USA}

\author[0000-0003-1197-0902]{ChangHoon~Hahn}
\affiliation{Department of Astrophysical Sciences, Peyton Hall, Princeton University, 4 Ivy Lane, Princeton, NJ 08544, USA}

\author[0000-0001-6941-8411]{Grayson~C.~Petter}
\affiliation{Department of Physics and Astronomy, Dartmouth College, 6127 Wilder Laboratory, Hanover, NH 03755, USA}

\author[0000-0003-1468-9526]{Ryan~C.~Hickox}
\affiliation{Department of Physics and Astronomy, Dartmouth College, 6127 Wilder Laboratory, Hanover, NH 03755, USA}

%\author{Friends}

\correspondingauthor{Rodrigo~C\'ordova~Rosado}
\email{rodrigoc@princeton.edu}

\begin{abstract}

An unresolved question in studies of active galactic nuclei (AGN) is whether their different classes probe different evolutionary stages of black hole--host galaxy interaction. We present the projected two-point cross-correlation function between a sample of Dark Energy Spectroscopic Instrument (DESI)-matched AGN selected from Hyper Suprime-Cam Subaru Strategic Program (HSC-SSP) optical + Wide-field Infrared Survey Explorer ($WISE$) mid-IR photometry, and DESI-designated luminous red galaxies, for $z\in 0.5-1.0$. The total overlap area is 43.4 deg$^2$, including $\sim27,000$ spectroscopic LRGs in our redshift range. We visually classified 1,991 matched HSC-DESI objects in our redshift range, spectroscopically confirming that 1,517 ($76\%$) of them are AGN. Of these 1,517 objects, $73\%$ are broad-line AGN, $27\%$ are obscured AGN. We infer that the parent HSC+$WISE$ AGN catalog has a number density of at least $\sim 240$ deg$^{-2}$, confirming it is one of the most complete optical/infrared AGN catalog to date. We investigate the AGN clustering as a function of the spectroscopic classification and infer the halo mass for each sample. The inferred average mass of the halos $\langle M_h\rangle$ that host unobscured broad-line AGN ($M_h \approx 10^{13.4}h^{-1}M_\odot$) is $\sim 5.5\times$ larger than the halos that host obscured AGN ($M_h \approx 10^{12.6}\, h^{-1}M_\odot$), at $2.8\sigma$ significance, in the same sense as our prior work based on photometric redshifts. This suggests that we may relax our concerns about systematic shifts in the inferred redshift distribution producing this halo mass difference. While we do not yet find statistically significant spectroscopic evidence that unobscured AGN reside in more massive halos than their obscured counterparts, further analyses are necessary to distinguish if more complex evolutionary histories are needed to model these AGN populations.

\end{abstract}


\section{Introduction} 

Active galactic nuclei (AGN) are the signposts of supermassive black hole (SMBH) growth in the Universe \citep{schmidt_3c_1963, kormendy_coevolution_2013}. By studying the galaxies that host these intense periods of matter accretion onto the SMBH, we learn about how these black holes influence the formation and evolution of galaxies \citep{kormendy_inward_1995, kormendy_coevolution_2013}. Understanding BH and galaxy co-evolution through the observation of active galaxies is critical to appreciate AGN feedback and the effect of SMBH's on the local environment and galaxy formation writ large \citep{fabian_observational_2012, kormendy_coevolution_2013, heckman_coevolution_2014}. 

This paper is concerned with the relationship between different AGN spectroscopic classes and how they may describe different evolutionary steps in galaxy formation \citep{hickox_host_2009}. Traditionally, AGN can be split into unobscured (Type I) and obscured (Type II) populations. The standard unification model proposes that all AGN are the same kind of object, with different inclinations relative to our line of sight \citep{antonucci_unified_1993, urry_unified_1995, netzer_revisiting_2015}. A dusty torus (dust beyond the sublimation radius at $\sim 1$ pc, with a relatively large covering fraction) screens the broadened emission from the sub-pc broad-line region around the accretion disk in obscured (Type II) objects, while in unobscured or Type I objects our line of sight does not intercept the torus \citep[see][for a recent review]{almeida_nuclear_2017}. Key diagnostics of AGN activity include the presence of broad permitted lines emerging from the region light-months around the black hole, and line ratios from high-energy excitation lines \citep[c.f.][]{baldwin_classification_1981, kewley_theoretical_2001, kewley_theoretical_2013, almeida_nuclear_2017, backhaus_clear_2022}. Classification tools to distinguish AGN as obscured vs. unobscured in the UV/optical include spectral slope (unobscured sources are typically blue, while obscured sources are galaxy-continuum--dominated), and the width of permitted emission lines.  

Previous studies have shown that there is some correlation between the degree of obscuration and possible AGN evolutionary stages \citep{sanders_ultraluminous_1988, canalizo_quasi-stellar_2001, hickox_clustering_2011, allevato_clustering_2014, fawcett_striking_2023}, suggesting that there is more to the story than just inclination. A number of results have also suggested that obscured AGN are more likely to be be undergoing a galaxy merger than their unobscured counterparts \citep{mihos_triggering_1994, mihos_gasdynamics_1996, blain_dust-obscured_1999, urrutia_evidence_2008, koss_merging_2010, ellison_galaxy_2011, ellison_galaxy_2013, glikman_major_2015, ellison_definitive_2019, goulding_galaxy_2018, secrest_x-ray_2020, ricci_hard_2021}. However, the relative importance of torus-scale and galaxy-scale dust obscuration is unclear \citep{goulding_towards_2009, goulding_deep_2012}. These studies show it is challenging to resolve the dichotomy of narrow-line (NL) and broad-line (BL) with just an inclination model, or if evolutionary tracks that produce different features near the site of accretion are also possible, as has been previously suggested \citep{hopkins_cosmological_2008, hickox_host_2009}.

An effective means of isolating global AGN properties is to infer the host dark matter (DM) halo mass via clustering techniques. This has been a preeminent tool to infer the ensemble properties of AGN populations and the galaxies that host them \citep{osmer_three-dimensional_1981, shaver_clustering_1984, shanks_spatial_1987, iovino_clustering_1988, andreani_evolution_1992, mo_quasar_1993, shanks_qso_1994, croom_qso_1996, la_franca_quasar_1998, martini_quasar_2001, croom_2df_2005, lidz_luminosity_2006, shen_biases_2008, toba_clustering_2017, he_clustering_2018,chaussidon_angular_2022, arita_subaru_2023}. A wide range of results have not led to a consensus on whether unobscured or obscured AGN are hosted in more massive halos, or if there is a difference at all \citep{hickox_clustering_2011, allevato_clustering_2014, dipompeo_angular_2014, dipompeo_updated_2016, jiang_differences_2016, dipompeo_characteristic_2017, koutoulidis_dependence_2018, powell_swiftbat_2018, petter_host_2023, li_black_2024}. With new multi-wavelength deep imaging and spectroscopic wide-field surveys, we have the tools and datasets to revisit this question, including deep optical imaging, a high number density sample of AGN, individual redshifts for each object, and now a complementing spectroscopic sample.

In \cite{cordova_rosado_cross-correlation_2024}, we sought to control for the many systematic effects that make an interpretation of halo mass differences from clustering measurements difficult. Using an unsupervised machine-learning selection of AGN from Hyper Suprime-Cam (HSC) optical and Wide-field Infrared Explorer (\textit{WISE}) mid-infrared (MIR) fluxes and colors prepared by Goulding et al. (in-prep.), we took our sample of $>28,000$ AGN (after applying luminosity and redshift thresholds), split them into unobscured and obscured subsamples, and measured the angular correlation function for each sub-type. We had a photometric redshift for each AGN in the sample (and $\sim 15\%$ had spectroscopic redshifts), which we used to build the $dN/dz$ used to construct the halo model to fit the clustering. Using a maximum likelihood analysis to estimate the average halo mass for each sample, we found that unobscured AGN reside in DM halos that are $\sim 5\times$ more massive than the halos in which obscured AGN reside, at $3.5\sigma$ statistical significance. While we had unprecedented statistical power, there was still the possibility of substantial systematic uncertainty in the photometric redshift distributions that could have mimicked a physical result. To ameliorate this point of discussion, we now turn to a complementary study with a fully spectroscopic sample.

In this work we again rely on the AGN selection by Goulding et al. (in-prep.), which contains the highest number density of AGN to date ($\gtrsim 340$ deg$^{-2}$) as estimated by initial follow-up spectroscopy \citep{hviding_spectroscopic_2024}. We match these objects with early release data from the Dark Energy Spectroscopic Instrument \citep[DESI, ][]{desi_collaboration_desi_2016}, and recover the spectra for $>4,000$ objects. We inspect the spectra of selected and matched objects, classifying them primarily as either BL or NL AGN. This verification step allows us to estimate the spectroscopically-confirmed AGN number density of the full Goulding et al. (in-prep.) catalog. Applying the same luminosity threshold and a similar redshift constraint, we calculate the projected two-point correlation function, which makes use of the information gained by having a spectroscopic redshift for each object in our analysis. We cross-correlate DESI LRGs and our DESI-matched AGN sample, and investigate the clustering amplitudes for the different AGN populations. Cross-correlations are less sensitive to systematic uncertainties that are not in common between the input datasets and provide increased S/N for a sparse sample like our AGN. 

This paper is organized as follows. In Section \ref{sec:data}, we summarize the datasets used in this analysis, including the matching procedure between HSC+\textit{WISE} AGN and DESI. In Section \ref{sec:methods}, we outline our methodologies for the machine classification of the AGN spectra, and define the two-point correlation function calculation, the uncertainty estimation, and parameter fitting and interpretation procedure. We detail the AGN classification results in Section \ref{sec:classification}, and our LRG autocorrelations and the cross-correlations with the AGN sub-samples in Section \ref{sec:clustering}. We discuss the implications of our results in Section \ref{sec:disc}, and conclude in Section \ref{sec:conclu}.

Throughout this analysis, we adopt a ``Planck 2018''  $\Lambda$CDM cosmology \citep{planck_collaboration_planck_2020}, with $h = H_0/100\,{\rm km\, s^{-1} Mpc^{-1}} = 0.67$, $\Omega_c = 0.1198/h^2 = 0.267$, $\Omega_b = 0.02233/h^2 = 0.0497$, $n_s = 0.9652$, and $\sigma_8 = 0.8101$. Quantities defined with a $\log$ are exclusively $\log_{10}$ values. Magnitudes are expressed in the AB system \citep{oke_secondary_1983}. In the context of galaxy bias and halo mass parametrization, we use the \cite{tinker_large-scale_2010} formalism with $\Delta = 200$ (the spherical overdensity radius definition). A correction for foreground dust extinction is applied to all observations as supplied in the HSC catalog \citep{aihara_third_2022} based on \cite{schlegel_maps_1998}.

\section{Data} \label{sec:data}

\begin{figure*}
    \centering
    \includegraphics[width = \linewidth]{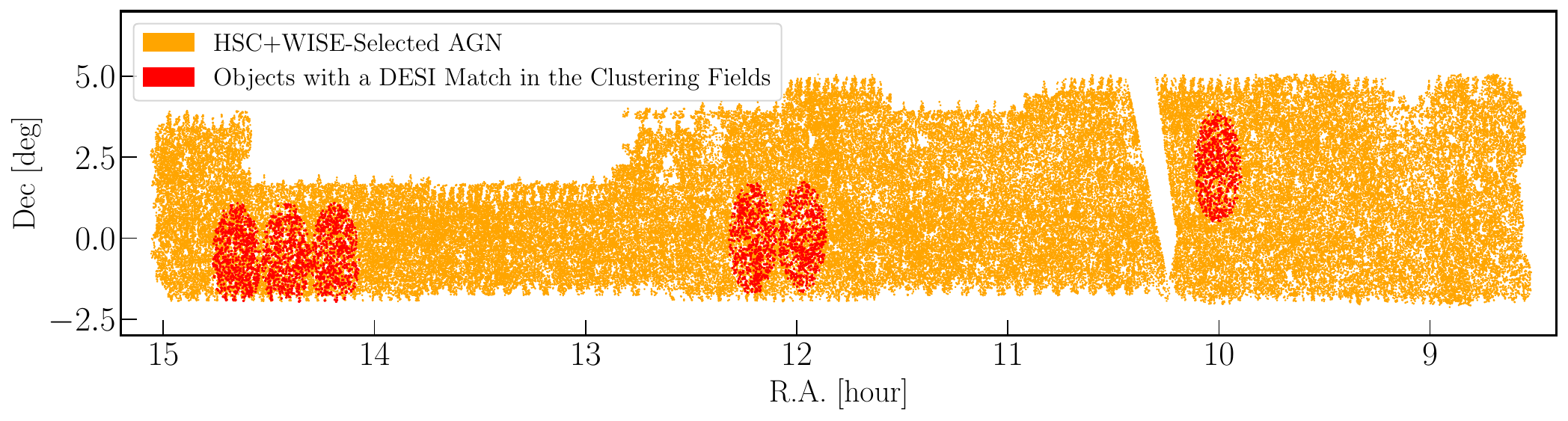}
    \caption{The HSC+WISE-selected AGN sample from Goulding et al. (in-prep.) for the HSC GAMA field, and its matches in the DESI EDR. The HSC GAMA field, centered at approximately at R.A., Dec. (11.7h, 0$^\circ$), includes all masking choices detailed in \S \ref{sec:masking}. Our procedure to select matches in the DESI EDR catalog is detailed in \S \ref{sec:matched_sample}. The three contiguous regions of overlap -- which we label GAMA10h, GAMA12h, and GAMA14h after their approximate hour coordinate in R.A. -- are treated independently in the analysis. The angular scales for the x- and y-axes are stretched inhomogeneously, making the circular DESI fields appear elongated.}
    \label{fig:overlap}
\end{figure*}

\begin{table*}
\centering
\caption{Field properties and number of matched objects in the three overlap fields in GAMA ($z\in 0.0-1.4$). }
\begin{tabular}{c|ccc}

\hline \hline  

&GAMA10h & GAMA12h&GAMA14h\\
\hline
Area [deg$^2$]  & 7.6  &  16.0 & 19.8 \\

$N_{obj}$ (DESI LRGs)  & 6,777 &  12,692  &16,216 \\

$N_{obj}$ (Matched AGN)  & 707 & 1,407 &  2,108 \\

$N_{obj}$ (Matched unobscured AGN)  & 225 & 470  & 832  \\

$N_{obj}$ (Matched reddened AGN)  & 245 & 471 & 510 \\

$N_{obj}$ (Matched obscured AGN)  & 223 & 429  &697 \\ 

\hline \hline 

\end{tabular} \label{tab:matched}

\end{table*}

\subsection{HSC Survey}

Hyper Suprime-Cam is a wide-field prime focus camera mounted on the 8.2\,m Subaru Telescope, atop Maunakea, Hawai'i \citep{miyazaki_hyper_2018}. The HSC - Subaru Strategic Program (SSP) \citep{aihara_hyper_2018} optimizes the 1.77\,deg$^2$ field of view by using 360 nights on Subaru to study galactic history out to $z\sim7$ with three observing strategies and \textit{grizy} wide-band filters \citep{kawanomoto_hyper_2018}. We point the reader to the most recent (PDR3) release of HSC-SSP data, described in \cite{aihara_third_2022}. We use the 670\,deg$^2$ (as of the release of PDR3) full-depth full-color Wide survey to obtain the properties of galaxies positioned across $\sim 2\%$ of the sky to implement an object cross match with the DESI, detailed below. Photometric redshifts (photo-$z$'s) are obtained from fits to the spectral energy distribution (SED) of each object, following the methods outlined in \cite{tanaka_photometric_2018} and the template-fitting algorithm \texttt{Mizuki} \citep{tanaka_photometric_2015}.

\subsection{Photometric Masking} \label{sec:masking}

We apply the HSC + \textit{WISE} masking strategy for the HSC fields as used by \cite{cordova_rosado_cross-correlation_2024}, and review salient details here. 
% Bright Star Masking
The HSC PDR3 have been released with a bright star and full-depth-full-color-mask, as described in \cite{coupon_bright-star_2018} and \cite{aihara_third_2022}. Using the Gaia DR2 bright star catalog \citep{gaia_collaboration_gaia_2018}, bright sources are identified and the affected sky regions are removed from the survey area. The source mask constructed based on the \textit{WISE} imaging data \citep{wright_wide-field_2010, cutri_vizier_2012} is detailed in Goulding et al. (in-prep.). Using \textit{WISE} catalog flags, we remove all objects with CCFLAGS=`H {\tt or} D {\tt or} X {\tt or} P'. The regions where these objects are found are affected by data issues. In addition, we also identify and remove regions where there is striping from uneven observation depth (such as $\geq 7 \sigma$ overdensities relative to the average across the field), moonlight contamination or additional artifacts, removing $ \sim 56$ deg$^2$.

\subsection{HSC and WISE-Selected Parent AGN Sample} \label{sec:AGNsample}

As we detailed in \cite{cordova_rosado_cross-correlation_2024}, our analysis relies on the accurate selection of AGN from the total galaxy sample obtained by HSC. There, we described the process for identifying and classifying objects as AGN based on their photometric properties. We refer the reader to \S 2.4 of that paper, and review the salient details here. 

Leveraging the deep optical photometry from HSC-SSP in combination with mid-IR photometry from the Wide-field Infrared Survey Explorer (\textit{WISE}) \citep{wright_wide-field_2010}, we select an AGN sample using unsupervised machine learning tools (hereafter referred to as the HSC+\textit{WISE} sample). This process has the benefit of using multiple components of the electromagnetic spectrum to find the AGN from $z\in0.1-3$, though we will focus on $z<1.5$ for our analyses. Full details of this method are explained in Goulding et al. (in prep.). With a maximum likelihood estimator, HSC \textit{grizy} photometry is matched to sources detected with S/N$>5$ in their W1 photometry in the all\textit{WISE} and un\textit{WISE} catalogs \citep{cutri_vizier_2012, schlafly_unwise_2019}. We combine the source catalogs, and require sources to have S/N $>4,2,2$ in their $g$, $W2$ and $W3$ respectively photometry for the complete AGN sample in hand. Utilizing the Uniform Manifold Approximation \& Projection (UMAP) algorithm to identify similar properties in the sample \citep{mcinnes_umap_2018}, we take in all the multi-dimensional color, magnitude, and source size information, and project these higher-dimensional correlations onto a two-dimensional representation. This results in an effective isolation of the AGN from the rest of the galaxy population. This paper will complement prior analyses by \cite{hviding_spectroscopic_2024}, who showed how these UMAP-selected AGN are consistently confirmed to be AGN via spectroscopic classification.

Goulding et al. (in-prep.) provide photometric redshifts for all of the UMAP-classified AGN without a spectrum prior to DESI's observations. Utilizing the full \textit{g} through \textit{W3} photometry, they train an augmented Random Forest (RF) algorithm to solve for each object's $p(z)$ distribution. From initial testing, these RF-based photometric redshifts perform equally well for both Type I and Type II AGN out to $z \sim 3$ with an average precision of $\delta z$/(1+$z$)$\sim 0.02$ and 0.03, respectively. We compare this photometric redshift (photo-$z$) with the DESI spectroscopic redshift (spec-$z$) for the matched objects in \S \ref{sec:matched_sample}.

The key features of the photometric classification are summarized here, noting that these categories will be compared with the final spectroscopic classification and are not used for the ultimate clustering sample splits. Using the $g-W3$ color and redshift information from each object in the HSC AGN sample, Goulding et al. (in-prep) identify the distribution minima in this color-redshift space using a spectroscopically-trained K-Nearest Neighbor (KNN) algorithm to probabilistically label AGN as unobscured, reddened, or obscured. The unobscured and reddened categories are trained with broad-line (Type I) AGN spectra, with the reddened objects having more obscuration, while the obscured category was trained with heavily obscured narrow-line (Type II) AGN. These categories serve as a benchmark to understand how well photometric classifications match spectroscopic classifications from DESI observations, to which we now turn.

\subsection{DESI Spectroscopic Observations}

\begin{figure*}
    \centering
    \includegraphics[width = 0.95 \linewidth]{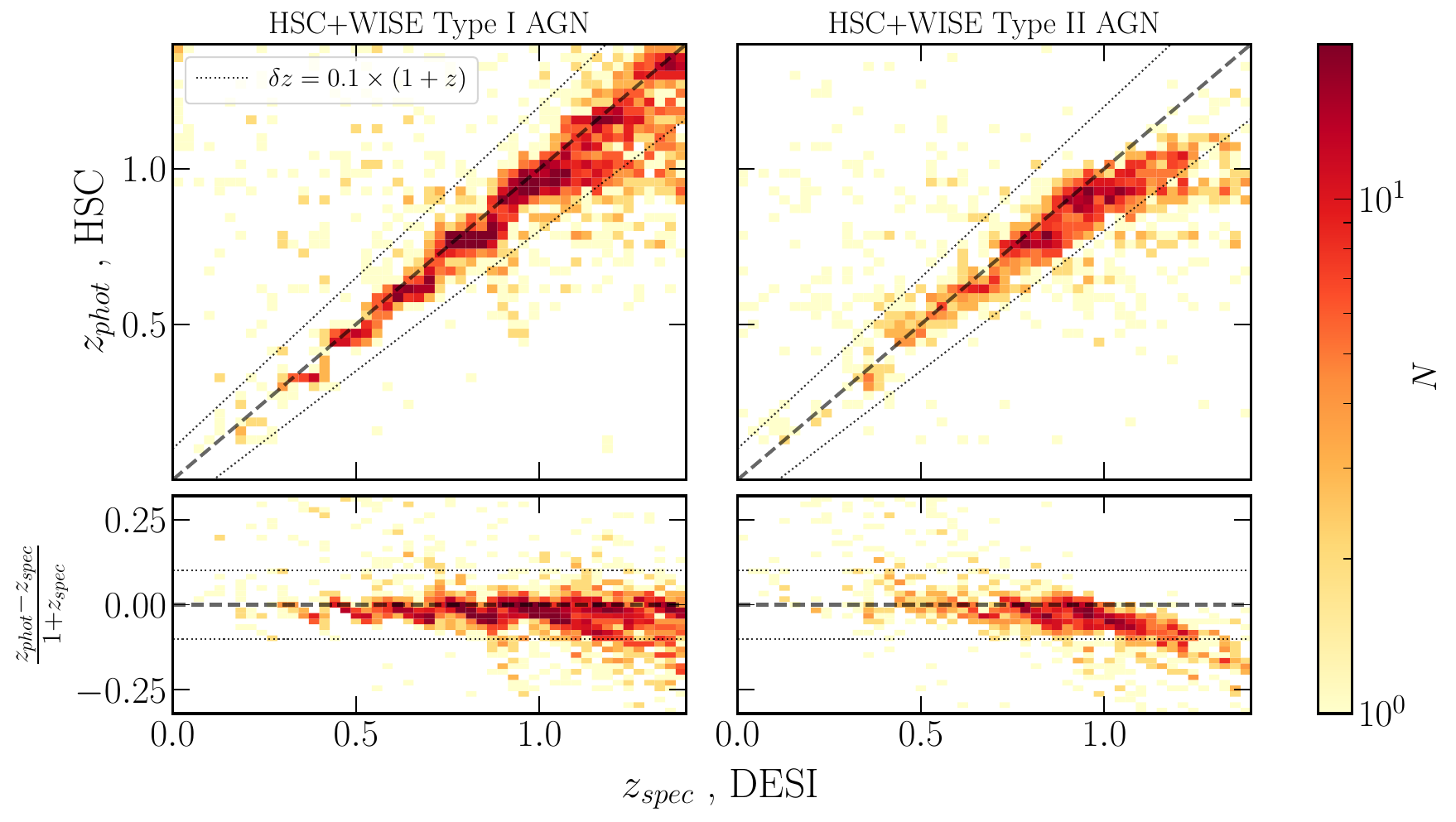}
    \caption{Redshift comparison for all DESI-matched HSC+\textit{WISE} AGN which did not already have a spectroscopic redshift, as a 2-D histogram colored by number of objects per cell $N$. We compare the photometric redshift from the HSC+\textit{WISE} catalog (described in \S \ref{sec:AGNsample}) with the DESI-provided spectroscopic redshift for all objects with $z<1.4$. We overlay a reference 1:1 relation as a black dashed line, and an uncertainty envelope of $\delta z = 0.1 \times (1+z)$ as a pair of dotted lines. We split the matched sample into Type I (unobscured and reddened) and Type II (obscured) AGN from the photometric classification we use in \cite{cordova_rosado_cross-correlation_2024} and reference in \S \ref{sec:AGNsample}. We note that the Type I AGN are more numerous, and their redshifts follow the 1:1 line with a MAD in their difference of $\sim0.05$, while the Type II AGN have a tilt of approximately $\delta z = 0.05 \times (1+z)$ and the MAD in their difference of $\sim0.07$. The HSC photo-$z$'s slightly overestimate the spectroscopic redshift, with the effect being most dramatic at $z>1.0$. The bottom panels show the difference between redshift estimates represented by $(z_{phot} - z_{spec})/(1+z_{spec})$.}
    \label{fig:AGN_redshift_comp}
\end{figure*}

\begin{figure*}
    \centering
    \includegraphics[width =  0.8\linewidth]{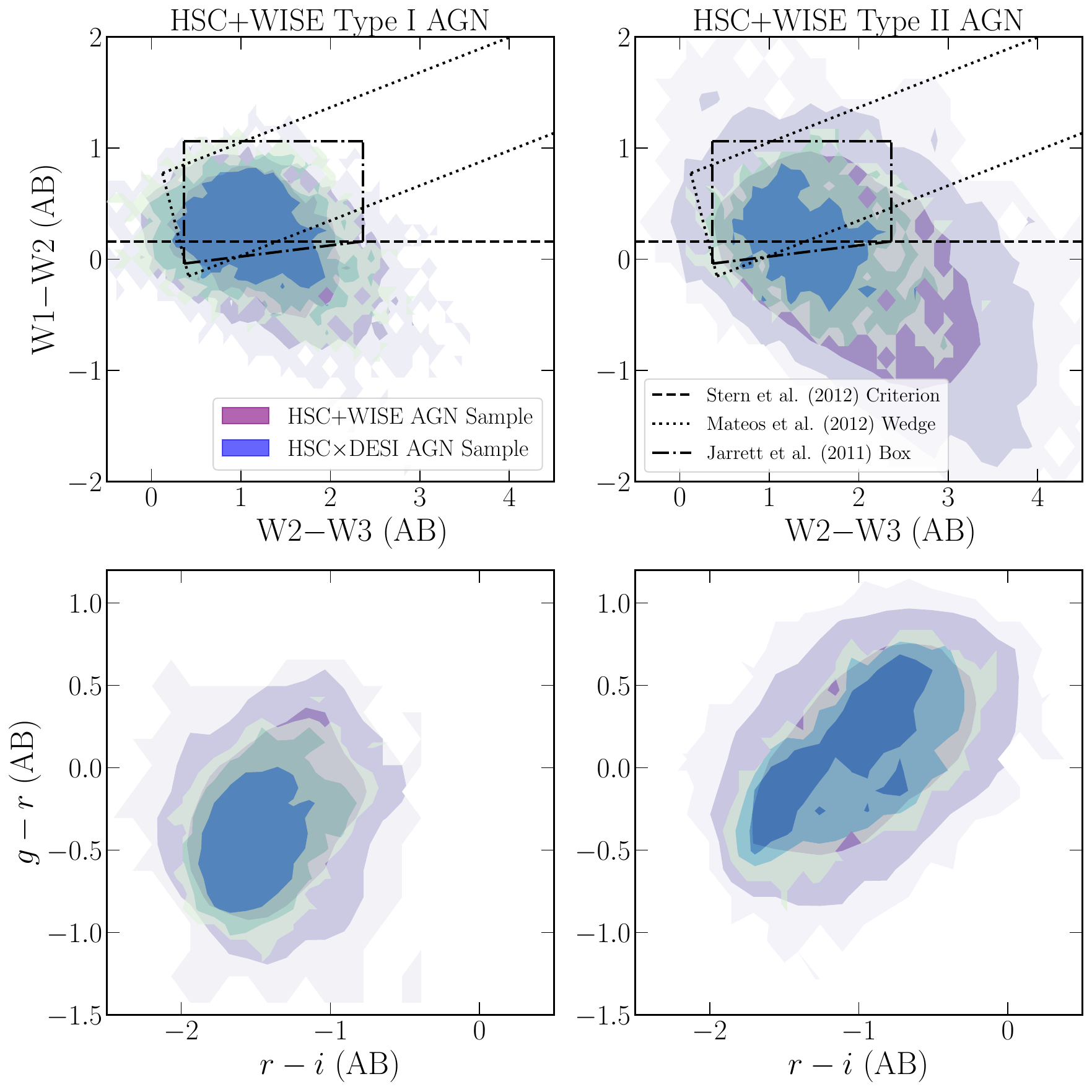}
    \caption{\textit{WISE} and optical photometric color-color distribution for the parent HSC+\textit{WISE} AGN catalog and the objects with a match in DESI. As in Figure \ref{fig:AGN_redshift_comp}, we split the HSC+\textit{WISE} AGN catalog using their photometric classifications into Type I (unobscured and reddened, left column) and Type II (obscured, right column). \textit{Top row:} the MIR W2$-$W3 vs. W1$-$W2 color space of our AGN. We overlay the \cite{stern_mid-infrared_2012} AGN selection threshold (W1$-$W2 $= 0.16$ in AB magnitudes) as a dashed line, the AB magnitude-converted \cite{mateos_using_2012} wedge in dotted lines, and the \cite{jarrett_spitzer-wise_2011} box in dot-dashed lines. Our unsupervised ML-selected AGN sample includes a significant number of objects that fall outside these canonical selection choices. The matched HSC$\times$DESI sample follows the color distribution of the parent HSC+\textit{WISE} sample, although the higher DESI S/N leading to an apparent undersampling of the bottom right color space for the Type II AGN, discussed in \S \ref{sec:matched_sample}. \textit{Bottom row:} the optical $r-i$ vs. $g-r$ color space of our AGN. The color distribution of the matched sample follows the contours of the parent sample closely, as expected given the similar optical selection constraints between the HSC+\textit{WISE} catalog and DESI.}
    \label{fig:wisecolorcomp}
\end{figure*}

\begin{figure*}
    \centering
    \includegraphics[width=0.8\linewidth]{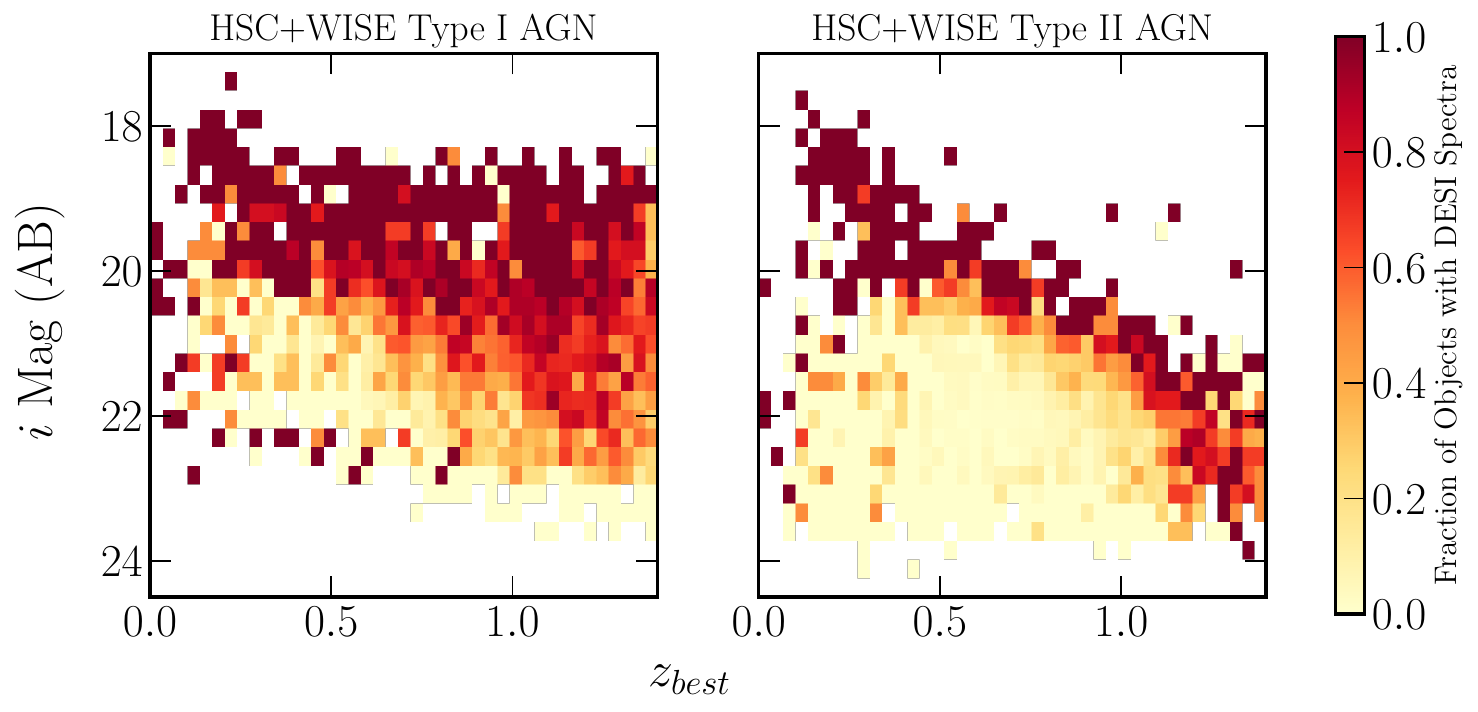}
    \caption{2-D distribution of redshift vs. $i$-band magnitude for all the HSC+\textit{WISE} objects in the HSC-DESI overlap fields in GAMA defined in \S \ref{sec:matched_sample}, colored by the fraction of objects that were spectroscopically observed by DESI. We again split the sample by the photometric AGN classification described in \S \ref{sec:AGNsample} in to Type I (left panel) Type II AGN (right panel). From the fraction of DESI observed-objects as a function of brightness and redshift, we observe the characteristic DESI S/N threshold as a function of redshift that selects for the brightest objects available. Thus, DESI poorly samples from the obscured and faint population of AGN in the HSC+\textit{WISE} catalog, and is much more complete for the characteristically brighter Type I AGN.}
    \label{fig:zvsImag}
\end{figure*}

DESI \citep{desi_collaboration_desi_2016} is a wide-field spectroscopic survey with a five-year goal to observe 14,000 deg$^2$ \citep[see][for an overview of the intrument]{desi_collaboration_overview_2022}. The DESI survey is performed on NOIRLab’s 4-m Mayall telescope at Kitt Peak National Observatory on the Tohono O'odham Nation in Arizona. 5,020 robotically-controlled fiber positioners, each holding a single optical fiber with $1.5\arcsec$ coverage, are installed on the 0.8-meter focal plane. 5,000 of these fibers direct light to one of 10 spectrographs, while 20 are dedicated to monitoring sky brightness. The total observed wavelength coverage is from $3600$ to $9824 \, \AA$. DESI has provided an Early Data Release (EDR) covering $1489$ deg$^2$ as part of its data validation process, including spectroscopic redshifts for all objects using their \texttt{RedRock} spec-$z$ fitting pipeline \citep{desi_collaboration_early_2023, desi_collaboration_validation_2024}. From these observations, several pointings overlap with the HSC survey, particularly in the GAMA field (see Figure \ref{fig:overlap}). We will build our analysis sample by matching HSC+\textit{WISE} AGN with DESI spectra, detailed below.

\subsubsection{DESI LRGs}\label{sec:lrgs}

In order to maximize the resolution of our correlation analysis, we perform a cross-correlation between a luminous red galaxy (LRG) sample and our AGN. From the complete EDR catalog, we isolate the DESI spectroscopic LRG sample via its unique survey target mask bit \citep{desi_collaboration_early_2023}. These objects are chosen in the DESI survey using DESI Legacy Survey \textit{g, r, z,} and \textit{WISE} \textit{W1} photometry to identify an $r < 21.5 (AB)$ elliptical galaxies' rest-optical obscured and low-MIR flux SED, as detailed in \cite{zhou_preliminary_2020}. We additionally verify there are no duplicates in our clustering catalog from repeat DESI observations. For the overlapping areas between DESI and HSC in the GAMA field, (Figure \ref{fig:overlap}), we identify 35,625 LRGs from the DESI sample, across redshifts $z\in 0.0-1.4$. This is split among three fields of overlap we have labeled GAMA10h, GAMA12h, and GAMA14h after their approximate R.A. coordinate, with a total area of $43.4$ deg$^2$. We will use only the subset of these objects that fall into our clustering analysis redshift bin (see \S \ref{sec:red}), in performing a projected two-point correlation function analysis (detailed in \S \ref{sec:corrfunc}).

\subsubsection{Matched HSC--DESI AGN}\label{sec:matched_sample}

We cross-match the full HSC AGN sample described in Goulding et al. (in-prep.) with the complete DESI EDR to identify all the possible spectra for the objects in our catalog. The selected overlap in the GAMA field is visualized in Figure \ref{fig:overlap}. We first perform an on-the-sky match between the parent HSC+\textit{WISE} AGN sample to the DESI EDR. We find 9,169 unique matches within a radius of $0.5\arcsec$ from each object's HSC-defined position.Hereafter, we will refer to this matched sample as the HSC$\times$DESI AGN. DESI's QSO selection is constrained with DESI Legacy Survey photometric values of $g - r < 1.3$ and $ r < 22.7$, having been chosen to match Type I AGN selections in SDSS \citep{yeche_preliminary_2020}. HSC and the DESI EDR overlap significantly in the GAMA and XMM fields, but we will only consider overlap areas with even observation depth in the GAMA field and objects at $z<1.4$ for our analysis. This leaves us with $4,222$ objects, split between the three overlap fields. These DESI-observed HSC+\textit{WISE} AGN and their photometric classification breakdown are shown in Table \ref{tab:matched}.

With these matched DESI spectra, we test the photometric redshift accuracy and completeness, and we assess whether the photometric classification is accurate using the optical spectral features and the \textit{WISE} luminosity. We calculate the rest-frame $L_{6 \mu m}$ luminosity for each AGN in the sample via the standard power law fitting to the MIR photometry from WISE, scaling the measured flux to a luminosity given the luminosity distance from the DESI spec-$z$. We compare the DESI spec-$z$ for all the matched HSC AGN with photo-$z$'s in Figure \ref{fig:AGN_redshift_comp}. We split the sample between the Type I (unobscured and reddened) and Type II (obscured) AGN. It illustrates the consistency in photo-$z$ measurement from the HSC and the spectroscopic DESI redshifts within an envelope of $dz/(1+z)=  0.1  $, though there is a $dz  \sim 0.1$ shift to lower photo-$z$ among the $z \gtrsim 1.0$ Type II AGN. We find a median absolute deviation (MAD) of the difference in redshift to be $\sim 0.05$ for the Type I AGN, and $\sim 0.07$ for the Type II AGN.

We also compare the HSC AGN sample with the DESI matches in color and magnitude space. We compare the \textit{WISE} MIR color distribution between the full AGN selected sample and the matched HSC$\times$DESI sample in the top row of Figure \ref{fig:wisecolorcomp}. We overlay the \cite{stern_mid-infrared_2012}, the \cite{mateos_using_2012} wedge, and the \cite{jarrett_spitzer-wise_2011} box for \textit{WISE} AGN selection, and note how a significant fraction of our parent and matched AGN catalog lie well below the limit and outside the wedge. As \cite{hviding_spectroscopic_2024} show, there are significant numbers of spectroscopically-confirmed AGN that lie below or outside these limits.  We note how the most reddened part of the (W2$-$W3 axis of) \textit{WISE} color-space is not as populated by the DESI matched-sources, indicative of the shallower magnitude limit for DESI sources. We compare the $r-i$ vs. $g-r$ color distribution in the bottom row of Figure \ref{fig:wisecolorcomp}, noting that the parent and matched catalogs distributions overlap in this optical color-color space. We visualize the difference in depth between HSC and DESI in Figure \ref{fig:zvsImag}. The right panel of Figure \ref{fig:zvsImag} shows how there is much shallower sampling of the obscured AGN population as a function of DESI's redshift-dependent S/N thresholds. There is a significant under-sampling of faint and obscured AGN by DESI (more so than the higher-flux Type I sample), which will limit our ability to compare our previous results with the obscured/Type II AGN identified in DESI. The HSC$\times$DESI sample will be the basis of our AGN classification analysis, investigating how well the photometric color vs. redshift inferred AGN categories map onto the optical/UV spectrum-determined type (based primarily on line widths and ratios).

\section{Methodology} \label{sec:methods}

\subsection{AGN Spectral Fitting with \texttt{PyQSOFit}}\label{sec:methodclass}

Given the overlap in the GAMA HSC fields with the DESI EDR, we spectroscopically analyze the $\sim 4,000$ DESI$\times$HSC AGN with $z_{spec} < 1.4$. A principal question is whether these unsupervised photometry and ML-selected objects possess optical spectroscopic signatures of AGN, and if so whether they are broad-line or narrow-line AGN. To do so, we fit each spectrum in this sample using the fitting tool \texttt{PyQSOFit} \citep{guo_pyqsofit_2018}, based on the \texttt{QSOFit} program developed by \cite{shen_qsofit_2019}. We will use this algorithm to perform a joint fit for a spectrum's emission lines and continuum. This includes models for the FeII pseudo-continuum, a power-law AGN continuum, and a host galaxy spectrum. The emission lines being fit are set by the user, with choices on the line width bounds, velocity offsets, and the number of Gaussians used to model the emission.

We now describe the parameter choices and components for the fitting procedure with \texttt{PyQSOFit}. Each spectrum is converted to restframe wavelengths before fitting, given the spectroscopic redshift provided by DESI. We bound the AGN continuum model as a power law $A \lambda^\beta$, with slope limits of $\beta \in [-5.0,  3.0]$. We fit the narrow and broad-lines with Gaussian profiles. Lines with broad and narrow components are simultaneously fit, but the width of each line is fit independently. We fit a single broad and a single narrow Gaussian model for each available permitted line (H$\alpha\, \lambda 6564$, H$\beta\, \lambda 4862$, H$\gamma\, \lambda 4340$, H$\delta\, \lambda 4102$, CIV$\, \lambda 1549$), with the exception of fitting the MgII line with a single broad-line component and two narrow-line components ($\lambda \lambda \, 2796, 2803$) to capture the doublet. There are strict bounds on the upper and lower velocity dispersion FWHM for narrow vs broad-line fits. The broad-lines are limited to FWHM$\in[400-7000]$ km/s, while the narrow-lines are constrained to be fit with profiles of FWHM$\in[10-300]$ km/s. We fit the available forbidden lines ([NII] $\lambda \lambda \, 6549, 6585$, [SII] $\lambda \lambda \, 6718, 6732$, [OIII] $\lambda \lambda \, 4959, 5007$, [NeIII]$\, \lambda 3870$, [OII] $ \lambda \lambda \, 3727, 3729$, [NeV]$\, \lambda 3426$) with narrow-line Gaussian profiles. We note that the spectral resolution of the DESI spectrograph is approximately $R\sim4000 \, (75 \,{\rm km/s})$, at the wavelength of [OII] at $z = 0.75$. We allow the line centers to be offset from the expected rest-frame value by up to $\pm50 \,\AA$.

The following are further details on the continuum fit. We use the the iron templates provided by \cite{pandey_new_2024} (using the non-turbulent template with hydrogen-ionizing photons flux $\log \Phi \,({\rm cm^{-2} s^{-1}})=17$ and gas density $\log n_h \,({\rm cm^{-3}})=12$), and constrain the Gaussian smoothing kernel bounds in the fit to be limited to $[10-500]$ km/s FWHM. We do not include a Balmer template in the fit, and allow galaxy host templates provided by \texttt{PyQSOFit} to be included where the minimization prefers them. Prior to any host subtraction, we fit for an initial continuum slope power law ($A \lambda^\beta$) to bound the observed slope $\beta$ for comparison with our prior $g-W3$ color-based AGN classifications. 

We apply this fitting procedure to all $z < 1.4$ DESI$\times$HSC AGN, and next sort the AGN into different spectroscopic categories based on the results. We detail the classification steps and decisions in \S \ref{sec:classification}, where we use the measurement and significance of these line profile fits to determine if each object observed by DESI is or is not an AGN, and if it can be classified as a broad-line (Type I) or narrow-line (Type II) AGN based on the optical spectrum.

\subsection{Clustering Measurement}\label{sec:corrfunc}

\subsubsection{The Projected Two-point Correlation Function}
We follow \cite{peebles_large-scale_1980} and \cite{davis_survey_1983} in defining the two-point correlation function $\xi(r)$. It measures the excess probability of finding galaxy pairs relative to a uniform random distribution as a function of the galaxy pair separation $r$. A limitation of this measurement, however, is the effect of Redshift Space Distortions (RSD) caused by the peculiar velocities of galaxies. To account for this, the correlation function can be estimated as a function of parameters $r_p$ and $\pi$, the galaxy separations perpendicular and parallel to the line of sight \citep{davis_survey_1983}. We use the standard (line of sight) projected two-point correlation function
\begin{equation}
    w_p (r_p) = 2 \int_0^{\pi_{\rm max}} \xi(r_p, \pi) d\pi,
\end{equation}
where $\xi(r_p, \pi)$ is measured correlation function from pair counting using the LS estimator \citep{landy_bias_1993}:
\begin{equation}
    \xi(r_p, \pi) = \frac{DD(r_p, \pi) - 2\,DR(r_p, \pi) + RR(r_p, \pi)}{RR(r_p, \pi)}.
\end{equation}
The normalized operator $XY(r_p, \pi)$ counts the number of pairs in bins of $r_p$ and $\pi$. The random points are generated such that they match the survey footprint of the data being correlated. After smoothing the $dN/dz$ of each dataset being correlated, we sample redshifts from this distribution to assign redshifts to the randoms, matching the input $z$ distribution. Thereafter, the  $\xi(r_p, \pi)$ measurement is integrated up to a chosen $\pi_{\rm max}$. The $\pi_{\rm max}$ is chosen to be the minimum value at which the $w_p(r_p)$ measurement has converged. Following \cite{allevato_xmm-newton_2011}, we iterate over $w_p (r_p, \pi_{max})$ tests and find that $\pi_{\rm max} = 60\, h^{-1}\, {\rm Mpc}$ gives adequate convergence and that minimizes introduced noise.

Using the \texttt{Corrfunc} implementation of $w_p (r_p)$ \citep{sinha_corrfunc_2020}, we calculate the clustering statistic over 12 logarithmically spaced bins from $0.1 - 40 \, h^{-1} \,{\rm Mpc}$.  We fit the measured correlation function with the model described in \S \ref{sec:model_def}.

\subsubsection{Uncertainty Estimation}
We use a jackknife procedure to calculate the total statistical and systematic error in a given per-field analysis. We split each of the three overlap fields into 25 equal-area regions ($0.3 - 0.8$ deg$^2$, depending on the field), and calculate the clustering signal from a set of 24 regions, excluding one region per iteration until all 25 have been removed once. We calculate the covariance matrix for each correlation function as defined in \cite{norberg_statistical_2009}:
\begin{equation}
    C_{jk} (x_i, x_j) = \frac{N-1}{N} \sum^N_{k = 1} (x^k_i - \bar{x}_i)  (x^k_j - \bar{x}_j),
\end{equation}
where $x_i$ is the $i$th measure of the statistic, i.e. $w_p(r_p)$, out of a total of $N =25$ measurements, and 
\begin{equation}
    \bar{x}_i = \sum^N_{k = 1} x^k_i / N .
\end{equation}

The full covariance matrix (for the chosen physical $r_p$ scales that we will fit) will be used to calculate the best fit. Additionally, the systematic uncertainty as a function of any field-to-field variability is constrained by comparing the recovered bias ($b$) and halo mass ($M_h$) values from each field (see \S \ref{sec:model_def}). We will also combine the bias and halo mass values from all fields to estimate the average and uncertainty from the complete sample with an inverse-variance weighted mean. The $1\sigma$ per-bin uncertainties represented in the illustrated correlation functions are drawn from the square root of the diagonal of each measured $w_p(r_p)$'s covariance matrix. 

\subsection{Clustering Interpretation}\label{sec:model_def}
\begin{figure*}
    \centering
    \includegraphics[width = \linewidth]{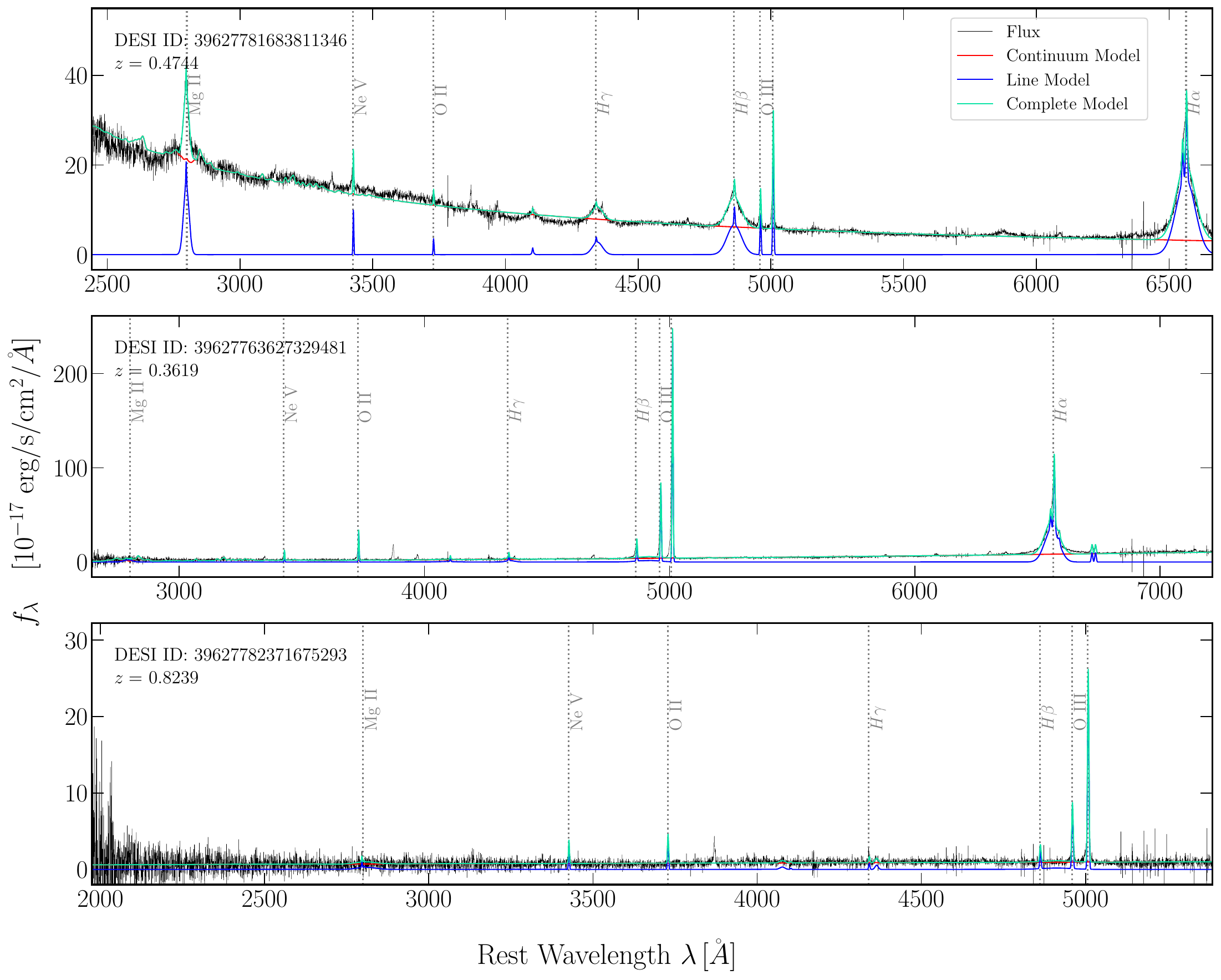}
    \caption{%Images of Spectra and fits
    Example HSC$\times$DESI AGN spectra and \texttt{PyQSOFit} fits to the continua and emission lines, following the procedure described in \S \ref{sec:methodclass}. The three spectra are chosen to be representative of a unobscured ($\beta \leq -1$) BL AGN (\textit{top}), a reddened ($\beta > -1$) BL AGN (\textit{middle}), and an obscured NL AGN (\textit{bottom}). The key emission lines we fit (as defined in \S \ref{sec:methodclass}) are highlighted with dotted vertical lines, where the spectra are presented in their rest-frame wavelength in units of $f_\lambda$. The DESI target ID's and redshift are shown in the upper left corner of each panel, and their R.A., Dec. coordinates are (from top to bottom): (178$^\circ$.6969,$-0^\circ$.1721), (182$^\circ$.3379,$-1^\circ$.1214), (219$^\circ$.6499,$-0^\circ$.1611).  The components to the spectral fits such as the continuum model and the emission line model are shown as red and blue lines, respectively. These showcase the impressive spectral resolution from DESI.}
    \label{fig:trip_spec}
\end{figure*}

We use a linear halo model to infer the clustering properties from the measured excess probability $w_p(r_p)$, based on peak background split theory \citep{sheth_large_1999}. We fit for the  linear bias term $b$ which describes the excess clustering signal from the galaxies relative to a linear dark matter halo clustering model. For our cross-correlations, the fitted multiplicative value is $b^2 = b_G b_A$, which is the product of the bias from each of the datasets (galaxies and AGN) being correlated. We estimate the galaxy bias from the autocorrelation of the LRG sample (i.e. $b^2 = b_G^2$), and thereafter use it to isolate the AGN bias ($b_A$) contribution in the cross term when computing the cross-correlation.

The DM correlation function can be modeled as a function of the matter power spectrum for the median redshift of the sample
\begin{equation}\label{eq:xir}
    \xi_{DM}(r) = \frac{1}{2\pi^2} \int P(k) \, k^2 \, j_0(kr)\,  dk,
\end{equation}
where $j_0(x)$ is the 0th order spherical Bessel function $\sin x/x$, and $P(k)$ is the matter power spectrum as a function of wavenumber $k$. We employ the $\xi(r, \bar{z})$ implementation in \texttt{CCL} to calculate our DM model %where $\bar{a}$ is the median scale factor from $a = 1/(1+z)$ 
\citep{chisari_core_2019}. The projected correlation function is thus
\begin{equation}
    w_{p, \,DM} (r_p) = 2 \int^\infty_{r_p} \frac{\xi_{DM}(r)}{\sqrt{r^2 - r_p^2}} \, r \, dr,
\end{equation}
keeping in mind the appropriate conversions such that $w_{p, \,DM} (r_p)$ is in units of $h^{-1} \, {\rm Mpc}$. 

With this DM-only model, we fit our measured $w_p (r_p)$ by scaling the model such that we minimize the single variable ($b^2$) log-likelihood

\begin{equation}\label{eq:logL}
    \log \mathcal{L} = \mathcal{R}^T \, C_{jk}^{-1} \, \mathcal{R},
\end{equation}
where $C_{jk}^{-1}$ is the inverse covariance matrix over the range of scales of interest, and the residual $\mathcal{R}$ is defined such that we fit for the $b^2$ parameter: 
\begin{equation}
    \mathcal{R} = w_p(r_p) - b^2 \,w_{p, \,DM} (r_p).
\end{equation}

We do not model the non-linear one-halo term, which would require a halo occupation distribution (HOD) treatment \citep[e.g.,][]{berlind_halo_2002}, and does not contribute to an estimate of the linear bias. We only fit over the scales $r_p > 1\,h^{-1}\,$Mpc in the measured correlation function, avoiding points on smaller physical scales in which the more complex HOD modeling would be necessary. Given these parameters and our 12 initial logarithmically-spaced angular bins, we now have a total of 7 data points from each correlation function to fit over, giving us 6 degrees of freedom (d.o.f.). We note that in fitting with the full covariance matrix over this physical range, the d.o.f. is not well defined given the off-diagonal correlations, and instead should be thought of as a heuristic description of the quality of the best fit. This method produces the standard bias analysis in the two-halo term clustering regime, and allows for an extension to estimate the mass of the average halo that produces the measured bias.

\subsubsection{Halo Mass Inference}

As in \cite{cordova_rosado_cross-correlation_2024}, we infer the average halo mass of our samples, $\langle M_h\rangle$, from the measured linear clustering bias given the above model and fitting algorithm. We use the \cite{tinker_large-scale_2010} parametrization of the halo mass function, and infer the halo masses following a similar procedure in  \cite{laurent_clustering_2017}. 

From the measured galaxy (AGN) bias, we calculate the lower halo mass threshold $M_{h,\,min}$ by solving for this value in
\begin{equation}
    b_1(z,M_{h, min}) = \frac{\int^{\infty}_{M_{h, min}}  \frac{dn}{dM} \langle N(M) \rangle b(z,M) dM }{\int^{\infty}_{M_{h, min}} \frac{dn}{dM} \langle N(M) \rangle dM} , 
\end{equation}
where $b_1$ is the measured clustering bias, $dn/dM$ is the halo mass function and the effective bias function $b(z,M)$ is as defined in \cite{tinker_large-scale_2010}, $\langle N(M) \rangle$ is the average halo occupation, which we assume is 1 for our analysis. 

We use this lower bound to estimate the average halo mass, $\langle M_h \rangle$, folding in the halo mass and redshift distribution of our sample:
\begin{equation}
    \langle M_{h} \rangle = \frac{\int^{\infty}_{M_{h,\,min}} \int_{z} M \frac{dn}{dM} \frac{dN}{dz} \langle N(M) \rangle\, dz \,dM  }{\int^{\infty}_{M_{h,\,min}} \int_{z} \frac{dn}{dM} \frac{dN}{dz} \langle N(M) \rangle\, dz \,dM } .
\end{equation}

This average halo mass estimate is representative of the underlying halo mass distribution that leads to the measured galaxy bias. By inferring the average host halo mass through this redshift and $dn/dM$ normalization, we more capture a portion of the cosmological distribution of halo masses in the Universe, rather than assuming all AGN are hosted in halos of a single representative mass $M_{\rm eff}$. As we note in \cite{cordova_rosado_cross-correlation_2024}, the absolute halo mass values are only comparable when treated with the same halo mass estimation formalism. We remind the reader that halo mass comparisons across analyses are complicated by subtle differences in the precise form of the bias-halo mass translation used. Using different prescriptions can introduce significant systematic shifts, such that relative halo mass values for different sub-samples are most useful for cross-analysis comparison.

\subsection{Redshift Bin}\label{sec:red}

In order to build sufficient statistics comparable to the results in our prior paper \citep{cordova_rosado_cross-correlation_2024}, we build a wide redshift bin of $0.5 < z < 1.0$. Herein we will calculate all $w_p(r_p)$ correlations, and investigate the clustering strength for the different AGN subtypes in our sample.

\begin{figure}
    \centering
    \includegraphics[width = 0.88\linewidth]{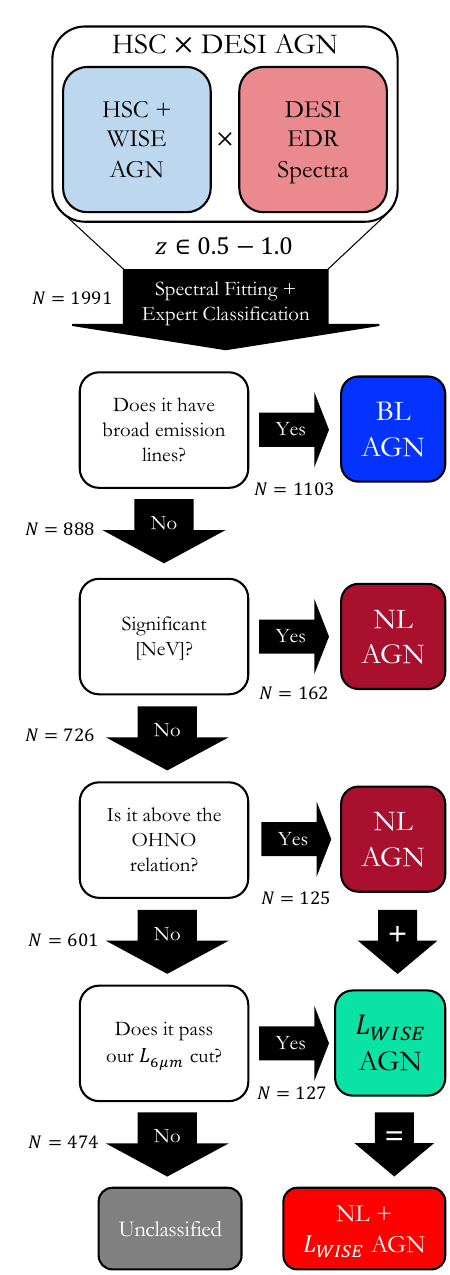}
    \caption{Diagram visually describing our spectroscopic classification. See \S \ref{sec:classification} for a detailed description of each of these steps.}
    \label{fig:flow}
\end{figure}

\begin{figure*}
    \centering
    \includegraphics[width=0.95\linewidth]{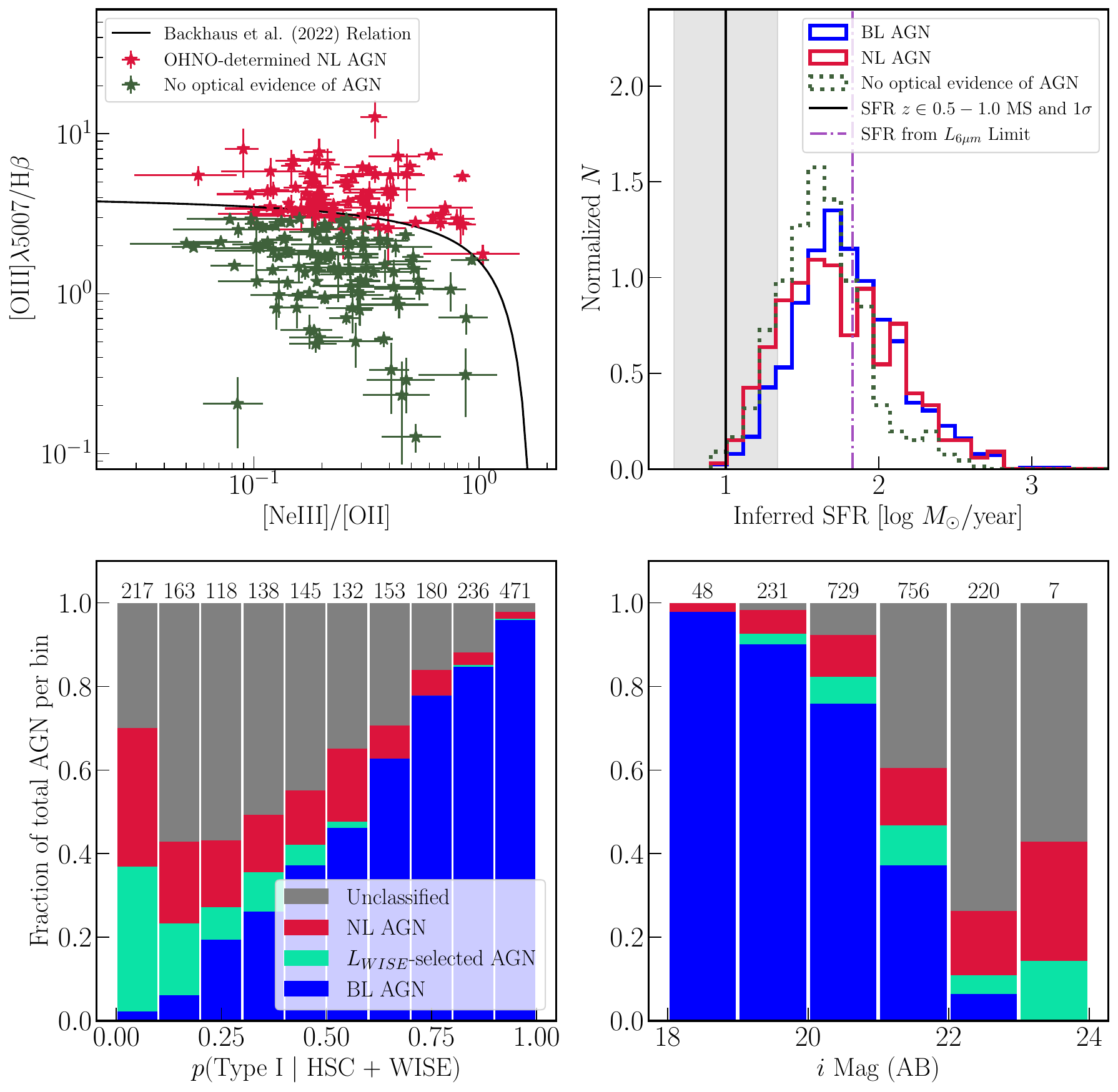}
    \caption{AGN spectral classification tests and outputs, detailed in \S \ref{sec:classification}. \textit{Top left:} the OHNO diagram \citep{backhaus_clear_2022} for AGN identification using line ratios available in this redshift and observed wavelength regime. \textit{Top right:} SFR distribution inferred by scaling each object's $L_{6\mu m}$ to $L_{24\mu m}$ with the ratio from a Sc galaxy template from \cite{polletta_spectral_2007}, and then scaling this luminosity to an SFR following the \cite{rieke_determining_2009} relation. We show the \cite{whitaker_star_2012} main sequence value and $1\sigma$ spread as a black line and shaded region to compare with our HSC$\times$DESI samples' distribution. \textit{Bottom left:} Sandplot representation of the distribution of the probability of the object being a Type I AGN given its HSC+\textit{WISE} photometry, split by final classification. The total number of objects in each bin is written above the bars. It shows a correlation between the abundance of the BL spectroscopic classification with the photometric probability they were Type I, and its converse for the NL objects. \textit{Bottom right:} $i$ magnitude sandplot distribution as a function of spectroscopic classification. As in the left panel, the total number in each bin is above the bars. We note how the brightest end of the distribution are dominated by BL AGN, while the dimmest objects are mostly NL AGN, $L_{WISE}$-selected AGN, and unclassified objects.}
    \label{fig:ohno}
\end{figure*}

\section{Results: AGN Classification} \label{sec:classification}

\subsection{Identification and Classification Process}
We detail our procedure for identifying and classifying AGN with DESI spectroscopy. As described in \S \ref{sec:methodclass}, we use \texttt{PyQSOFit} to fit the continua and emission lines for all 4,222 objects in the $z<1.4$ HSC$\times$DESI sample, i.e. HSC+\textit{WISE} photometrically-identified AGN with a matched DESI spectrum. Examples of these fits for characteristic unobscured BL, reddened BL, and NL AGN are shown in Figure \ref{fig:trip_spec}. The emission line and continua parameters we fit for each spectrum are necessary for the AGN sub-type spectroscopic classification.

We decided to visually inspect all objects for signs of broad-lines to corroborate and inspect any (lack of) significant broad-lines from \texttt{PyQSOFit}, or other issues like misattributed redshifts. Our classification procedure is summarized in the diagram in Figure \ref{fig:flow}.

The visual inspection was performed by three of the authors (RCR, ADG, and JEG), each examining the spectrum and individual line widths and fluxes for the $H\beta$ and MgII lines. Each of the classifiers indicated  -- for each object -- if it had clear signs of being a BL AGN, if it was a spurious BL fit to continua/had no BL emission, or if there was a redshift error. If any classifier identified the object as having an incorrect redshift from DESI's fitting, this object was removed from the analysis sample. We removed redshift failures ($\sim 2\%$ of the $z\in0.5-1.0$ sample, 43 objects out of 2,034) such that our final spectroscopically matched catalog of HSC$\times$DESI AGN has 1,991 objects in our clustering analysis redshift bin. The inspectors agreed on the BL AGN identification for $99.8\%$ of the objects, and when we did not we defaulted to labeling them as not having optical evidence of BL emission. There were no instances in which the three classifiers all disagreed with each other. We visually confirm and classify 1,103 of the objects as BL AGN based on the DESI optical spectrum. We find that $97\%$ of the objects we sorted into the BL category had \texttt{PyQSOFit}-determined $5\sigma$ non-zero detections of either MgII or H$\beta$ BL emission.

We sort the remaining 888 objects (of the total 1,991) into those having signs of AGN activity (or not) based on their line strengths and ratios from \texttt{PyQSOFit}. Our NL AGN selection has two steps. First, a ($\geq 3\sigma$) non-zero detection of the [NeV]$\, \lambda 3426$ line is taken as unambiguous evidence of an AGN \citep{schmitt_difference_1998, gilli_x-ray_2010, mignoli_obscured_2013}. Second, if [NeV] is not detected, we use the fluxes from the \texttt{PyQSOFit} line parameters to examine specific line ratios. We identify 162 objects with a significant [NeV] line, and classify them as NL AGN. Thereafter, following the parameterization in \cite{backhaus_clear_2022}, we place the remaining 726 unclassified objects on the OHNO diagram ([NeIII]/[OII] vs. [OIII] $\lambda \,5007$/ H$\beta$) so long as they had $3\sigma$ significant detections in all four lines. We find that 125 are consistent with or above the relation (see the top left panel of Figure \ref{fig:ohno}). We corroborate the \texttt{PyQSOFit}-derived fluxes for each line by re-fitting each continuum-subtracted $f_\lambda$ with an unbounded Gaussian profile -- as well as integrating the fluxes directly -- finding that the objects are sorted consistently across approaches. We label these objects as NL, or Type II, AGN. We note that this diagram is less reliable than other strong-line diagnostics \citep[c.f.][]{backhaus_clear_2022, scholtz_jades_2023, treiber_uncovering_2024}, but we have all the necessary lines for this diagnostic in our redshift range. We have a total of 287 NL AGN, out of the total 888 objects we have considered for NL classification. The remaining 601 objects do not display any optical evidence of AGN activity, some from not passing the line ratio tests and others from not having significant line emission, and we consider if the \textit{WISE} MIR photometry provides any additional information.

\subsection{AGN Selection from MIR Luminosity}

An essential component of the unsupervised machine learning identification of HSC sources as AGN was the inclusion of the \textit{WISE} MIR photometry, which we use alongside the spec-$z$ to infer a MIR luminosity. MIR emission can arise from both an AGN and star formation, and knowing the luminosity in this wavelength range allows for a consideration of which of these two mechanisms is more likely or dominates.

We consider what the implied star formation rate (SFR) would be for these objects if we were to assume a conservative (relatively small $L_{24\mu m }$ to $L_{6\mu m }$ ratio) dust temperature galaxy SED, i.e. an Sc galaxy template from \cite{polletta_spectral_2007}. Using the SFR relation from Equation 9 of \cite{rieke_determining_2009}, we show the distribution of inferred SFR from our HSC$\times$DESI AGN samples in the top right panel of Figure \ref{fig:ohno}. We overlay the $M_\star \sim 10^{10} M_\odot$ star-forming main sequence and $1\sigma$ width of galaxy SFR for $z\in 0.5-1.0$ from \cite{whitaker_star_2012}, noting that virtually all of our objects lie above the median $10 M_\odot$ yr$^{-1}$ value. Following the above conservative scaling, the median value for the combined BL and NL sample would be $\sim 57 \, M_\sun$ yr$^{-1}$. Given that this value is above the $2\sigma$ upper bound of the \cite{whitaker_star_2012} main sequence, it suggests that a significant fraction of our objects have MIR luminosities that are not straightforwardly explained by star formation, but rather could be the sign of emission from an AGN. In our analysis below, we will impose the same $L_{6 \mu m } > 3 \times 10^{44}$ erg s$^{-1}$ threshold for our clustering sample as in \cite{cordova_rosado_cross-correlation_2024}, which corresponds to an SFR of $67 \, M_\sun$ yr$^{-1}$ (purple dash-dotted line in the top right panel of Figure \ref{fig:ohno}), under this conservative scaling. We use this threshold to select the objects in our sample with no optical evidence of AGN activity to define a $L_{WISE}$-selected AGN sample. This results in 127 $L_{WISE}$-selected AGN, and 474 unclassified objects from our parent HSC$\times$DESI sample of 1,991 objects for $z\in0.5-1.0$. If we use the SED scaling from an M82 starburst galaxy template, which infers higher SFR at a given MIR luminosity, we find that our $L_{6\mu m}$ threshold lies above the $5\sigma$ upper bound of the star formation main sequence at SFR$\sim691\, M_\sun$ yr$^{-1}$. Moreover, the entire inferred SFR distribution for our samples would lie above the $3\sigma$ upper bound of the star formation main sequence. We also note for the objects we label as unclassified that a significant fraction are likely AGN, given that upon stacking the DESI spectra of the upper $L_{6\mu m}$ $50\%$ of objects, we detect significant [NeV] emission. These results suggest that further observations are needed to understand these objects' emission mechanisms, and that many are possibly powered by AGN.

\subsection{Additional AGN Subsamples}
\begin{table*}[t]

\caption{Final expert classification results and clustering samples ($z\in 0.5-1.0$)}
\centering
\begin{tabular}{c|c|c|c|c|c|c}
\hline \hline  

\hline
& \multicolumn{2}{c|}{GAMA10h} & \multicolumn{2}{c|}{GAMA12h}&\multicolumn{2}{c}{GAMA14h}\\
\hline

DESI LRGs  & \multicolumn{2}{c|}{5,108}  & \multicolumn{2}{c|}{9,124}  & \multicolumn{2}{c}{12,708}   \\
\hline  

&Classified & $^{*}L_{6\mu m}$&Classified & $^{*}L_{6\mu m}$&Classified & $^{*}L_{6\mu m}$\\
\hline
All Objects                 & 338& 131 & 627 &251 &  1,026 &322 \\

Broad-line  AGN$^\dagger$              & 174& 81 & 339& 156 &  590 &230 \\

Narrow-line    AGN                & 53 &21 & 94 &44  & 140& 45  \\

$L_{WISE}$-Selected AGN &  29&29 & 51 &51 & 47 &47 \\

Unclassified Objects & 82 & -- & 143 & -- & 249 & -- \\

\hline \hline

\end{tabular}

\vspace{0.05in}
\raggedright

\footnotesize 
$^*$ Luminous AGN selection for clustering analysis ($L_{6\mu m} > 3\times 10^{44}$ erg s$^{-1}$) 

$^\dagger$ Visual inspection confirmation of broad-line emission for classification.

\label{tab:finalclass}
\end{table*}

\begin{figure}
    \centering
    \includegraphics[width = 0.91\linewidth]{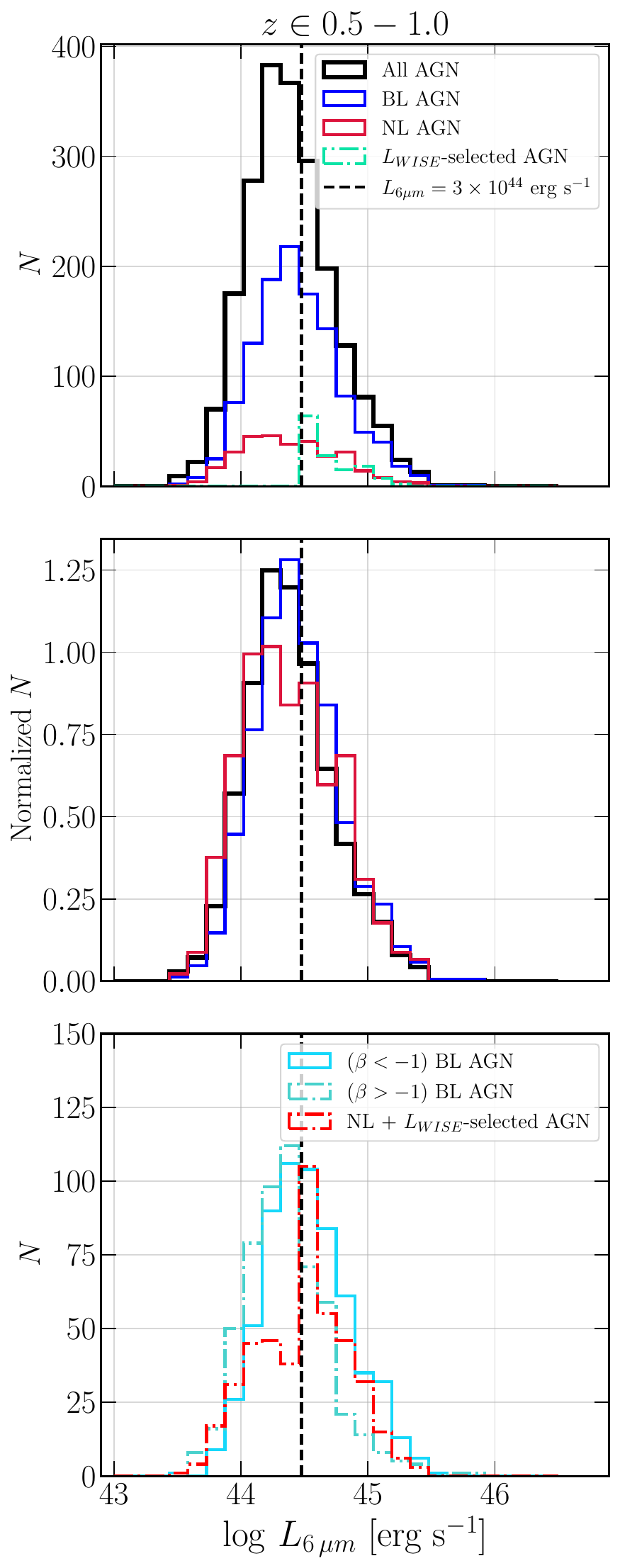}
    \caption{Luminosity histograms for each of our AGN (sub)samples in the $z\in0.5-1.0$ bin combining the three overlap fields in GAMA. The top and bottom panels show the original distributions, while the middle panel is normalized in order to show the consistency in the $L_{6\mu m}$ distribution between the complete, BL, and NL AGN samples. The vertical black dashed line is shared across the panels and denotes the $L_{6\mu m} = 3\times 10^{44}$ erg s$^-1$ lower limit for this analysis.}
    \label{fig:L6}
\end{figure}

\texttt{PyQSOFit}'s $A\lambda^\beta$ power law continuum fit to the pre-host subtracted spectrum allows us to differentiate the most unobscured BL AGN from the ``reddened'' AGN we examined in our prior work. Based on the distribution of measured $\beta$ values from the DESI-matched HSC+\textit{WISE} unobscured and reddened AGN, we select $\beta = -1$ as the appropriate discriminator between the unobscured $\beta \leq -1$ and reddened $\beta > -1$ spectroscopic BL AGN. See Figure \ref{fig:trip_spec} for example unobscured and reddened BL AGN. 

We also build an additional combination of the NL + $L_{WISE}$-selected AGN sample (the most analogous sample to the photometrically-defined obscured AGN in the HSC+\textit{WISE} catalog).

\subsection{Comparison with Photometric Classification}

With the spectroscopic classification described above, we test whether our prior HSC+\textit{WISE} photometric classifications were accurate. In the bottom left panel of Figure \ref{fig:ohno}, we show the fraction of BL, NL, and $L_{WISE}$-selected AGN as a function of their Type I AGN likelihood from the KNN algorithm classification used to sort the photometric HSC+\textit{WISE} AGN. We find that the photometric classification probability is well correlated with the spectroscopic classification. Conversely, the objects that were selected as having low chance of being Type I AGN (i.e. had a high probability of being Type II AGN) are spectroscopically classified as either NL, $L_{WISE}$-selected AGN, or they are unclassified. 

For consistency with our prior analysis in \cite{cordova_rosado_cross-correlation_2024}, we apply the same luminosity cut to our final clustering sample, such that all the considered AGN have $L_{6\mu m} > 3 \times 10^{44}$ erg s$^{-1}$. The luminosity distribution for each of our AGN subsamples is shown in Figure \ref{fig:L6}, along with the $L_{6 \mu m}$ threshold. We also show the normalized luminosity distribution in the middle panel of Figure \ref{fig:L6}, illustrating the consistency in the luminosity distribution for the different subsamples past the $L_{6\mu m}$ threshold. The resulting sample contains 704 AGN, 467 of which are BL objects, 110 NL, and the remaining 127 are $L_{WISE}$-selected AGN (split between the different overlap fields.). The results of this classification are shown in Table \ref{tab:finalclass}, tabulating all the objects that were visually classified and those that pass the $L_{6\mu m}$ threshold. The distribution of these classified objects in $i$-band magnitude are shown in the bottom right panel of Figure \ref{fig:ohno}, highlighting how the fraction of NL AGN, $L_{WISE}$-selected AGN, and unclassified objects dominate the fainter magnitudes.

This spectroscopic classification allows us to better understand the photometric classification that initially selects these AGN from the HSC and \textit{WISE} datasets. As \cite{hviding_spectroscopic_2024} previously showed, spectroscopic follow-up of these HSC+\textit{WISE} AGN probes how well the classification recovers BL as well as obscured and redenned objects, that previously were not significantly probed by surveys such as SDSS \citep{glikman_first-2mass_2012, assef_wise_2018, lyke_sloan_2020}. From the HSC$\times$DESI overlap sample, we find that we can confidently classify $76\%$ of them as AGN. From a total of 1,517 spectroscopically-confirmed AGN, $74\%$ are BL AGN, $19\%$ are NL AGN, and $9\%$ are $L_{WISE}$-selected AGN. The striking abundance of BL AGN in the spectroscopic sample is reflective of the DESI selection for blue \textit{WISE} colors and classic Type I quasars. The matched obscured (NL,  $L_{WISE}$-selected, and unclassified) objects in the HSC$\times$DESI sample are preferentially dimmer than their unobscured counterparts (see the bottom right panel of Figure \ref{fig:ohno}). This is coupled with the DESI magnitude limit ($r < 22.7 (AB)$ for their quasar selection, $r < 21.5 (AB)$ for LRGs) and S/N thresholds as a function of redshift (Figure \ref{fig:zvsImag}) and limits our current understanding of the complete Type II AGN sample probed by HSC+WISE. It presents an opportunity for additional observations in the future to constrain this population.

We will discuss the number density of these spectroscopically-classified objects and its implications for surveyed AGN populations in \S \ref{sec:disclass}. We now turn to the clustering measurements for these AGN sub-type samples.

\section{Results: Clustering Measurements}\label{sec:clustering}
\begin{figure}
    \centering
    \includegraphics[width = 1.\linewidth]{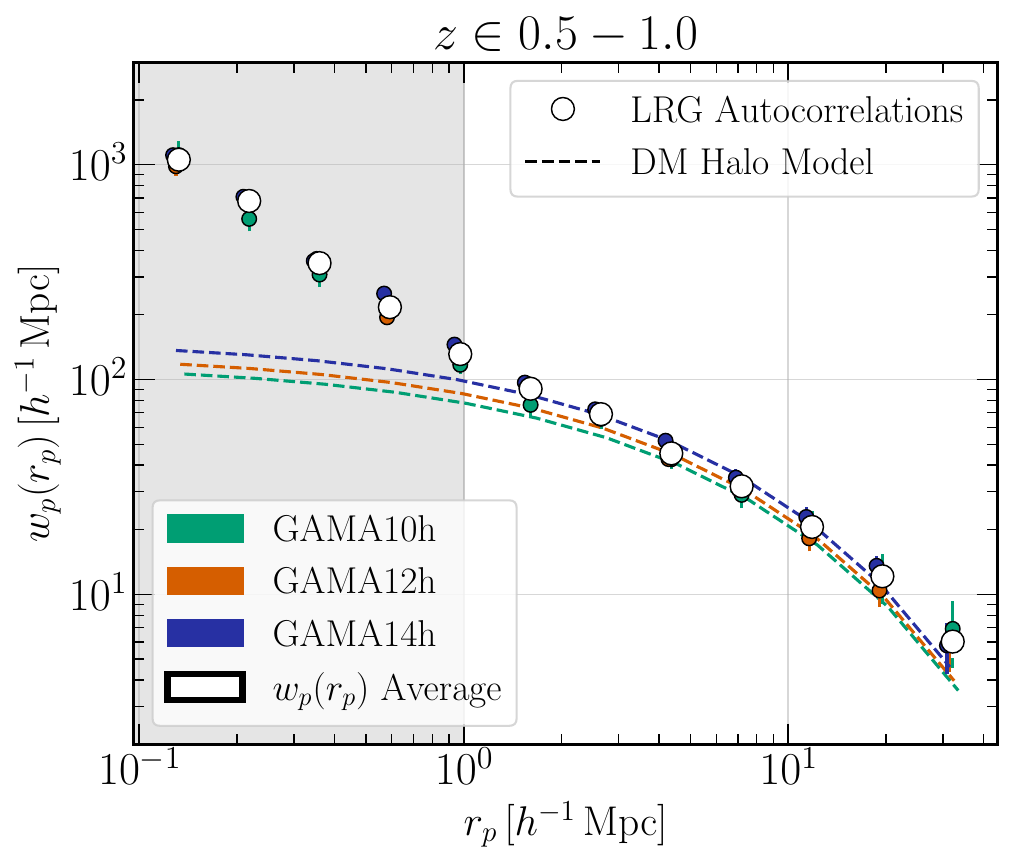}
    \caption{The measured DESI LRG projected two-point autocorrelation in our three HSC-DESI overlap fields for our $z\in0.5-1.0$ redshift bin. The open symbols are the per-bin inverse variance-weighted average correlation function across fields. The $1\sigma$ uncertainties are drawn from the square root of the diagonal of the jackknife covariance matrix for the sample. The dashed lines represent the fitted DM halo model to each field. We fit points for physical scales $r_p > 1\, h^{-1}{\rm Mpc}$, while the gray region contains the points at smaller scales excluded from the fit. Note that the model does not include the 1-halo term, which is why the data rise significantly above the model on small scales.}
    \label{fig:autocorr}
\end{figure}

\begin{figure*}
    \centering
    \includegraphics[width = 1.\linewidth]{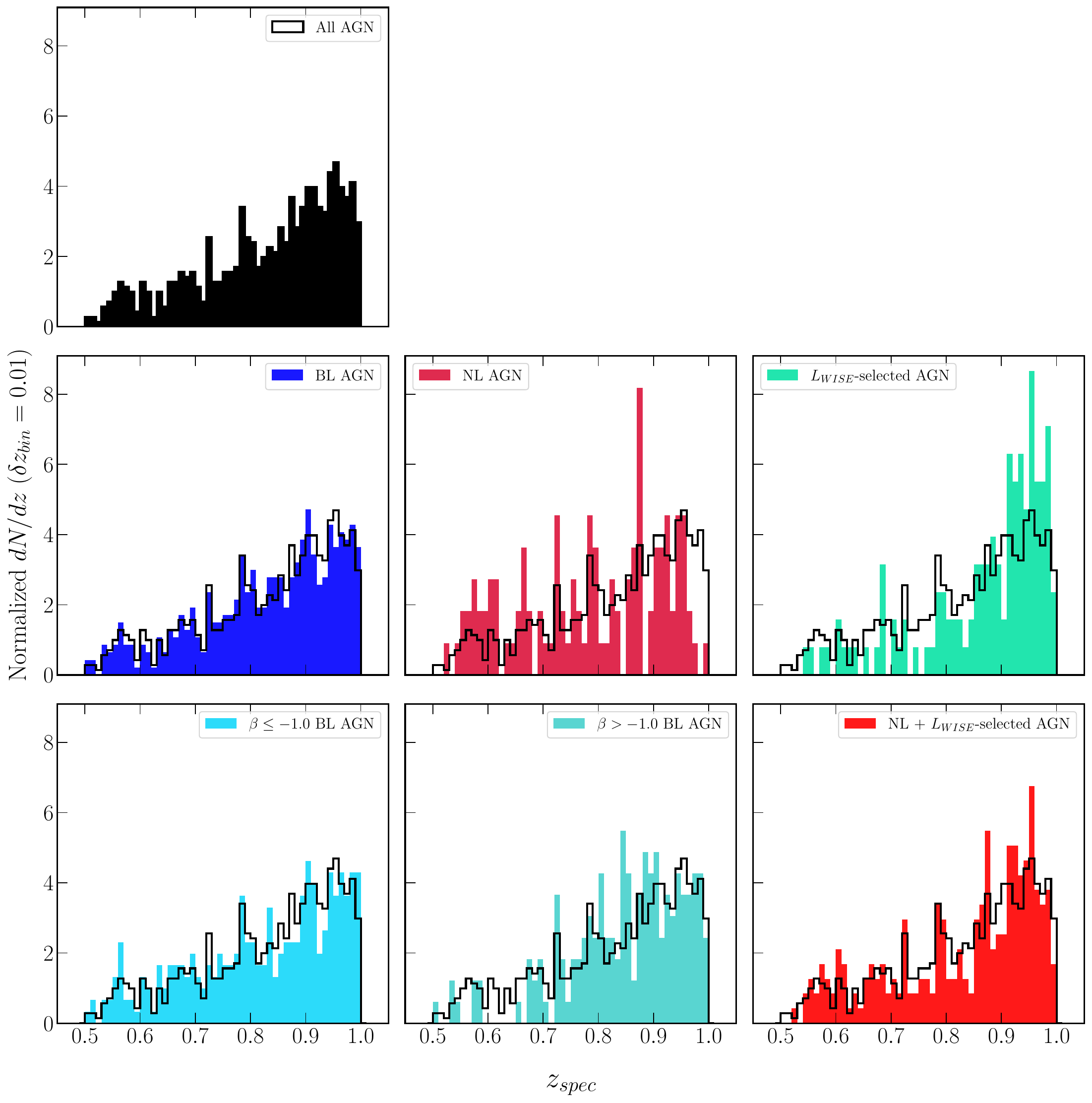}
    \caption{Normalized spectroscopic redshift distribution of each AGN (sub)sample for the clustering analysis when combining the catalogs from all GAMA overlap fields, after applying redshift and $L_{6\mu m}$ thresholds. The black line is shared among all panels to compare with the complete AGN sample's distribution.}
    \label{fig:dndz}
\end{figure*}

\begin{figure*}
    \centering
    \includegraphics[width = 1.\linewidth]{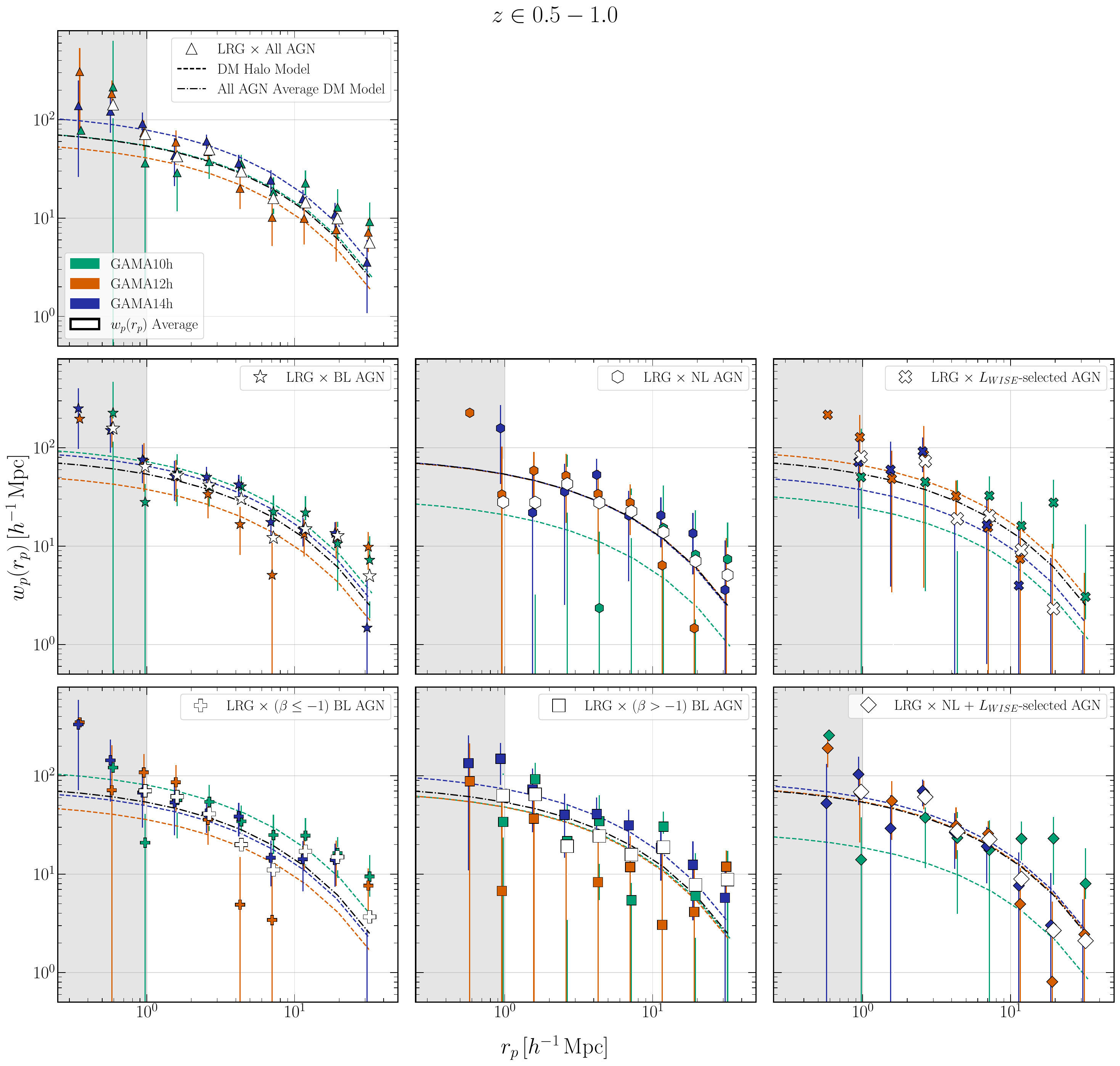}
    \caption{The projected two-point cross-correlation between DESI LRGs and our HSC$\times$DESI AGN (sub)samples for the three HSC-DESI overlap fields in the $z\in0.5-1.0$ redshift bin. The open symbols are the per-bin inverse variance-weighted average correlation function across fields. The $1\sigma$ uncertainties are taken from the square root of the diagonal of the jackknife covariance matrix for the sample. The dashed lines represent the fitted DM halo model to each field. We fit points for physical scales $r_p > 1\, h^{-1}{\rm Mpc}$, while the grey region contains the interior points excluded from the fit. We note the departure from the DM model on these small scales, as expected for the 1-halo term we do not model. The dot-dashed black line represents the full AGN sample average best-fit linear DM model, and is repeated across all panels as a reference line for ease of comparison. (From left to right, starting at the top) the cross-correlation of the full, BL, NL, and $L_{WISE}$-selected samples of AGN across the three overlap fields. Next, the cross-correlation of blue ($\beta \leq -1$) BL, and red ($\beta > -1$) BL and the NL + $L_{WISE}$-selected samples of AGN with the DESI LRGs. Symbols used for each (sub)sample here match those in Figure \ref{fig:masssumm}.}
    \label{fig:crosscorr}
\end{figure*}

\renewcommand{\arraystretch}{1.3}

\begin{table*}
\centering
\caption{Projected Two-point Correlation Function Fit Results} 
\begin{tabular}{ccccccc}
\hline \hline
\multirow{2}{*}{Subset} & \multirow{2}{*}{$N_{obj}$}  & $\langle L_{6 \mu m} \rangle$ & \multirow{2}{*}{$\langle z \rangle$} & \multirow{1}{*}{$\langle\chi^2\rangle$} & \multirow{2}{*}{$\langle b \rangle$} & $\langle M_h \rangle$ \\

 &  &   [$\log$ erg s$^{-1}$]  & & [6 d.o.f.]   & & [$\log h^{-1} M_\odot$] \\

\hline

LRGs                      & 26,940 & --  & $0.8\pm0.1$ & 19.0 & $2.3^{+0.1}_{-0.2}$ & -- \\
 
All AGN                    & 704    & $^* 44.7^{+0.3}_{-0.2}$  & $0.9^{+0.1}_{-0.2}$ & 7.4 & $1.7\pm0.2$ &  $ 13.1^{+0.1}_{-0.2}$ \\

BL AGN                     & 467   & $^* 44.7^{+0.3}_{-0.2}$ & $0.9^{+0.1}_{-0.2}$ & 6.0 & $1.8\pm0.2$ & $ 13.2\pm0.1$ \\

($\beta \leq -1$) BL AGN  & 303   & $^* 44.7^{+0.3}_{-0.2}$ & $0.9^{+0.1}_{-0.2}$ & 6.5 & $1.8\pm0.2$ & $ 13.4^{+0.1}_{-0.2}$ \\

($\beta > -1$) BL AGN     & 164 &   $^* 44.7^{+0.3}_{-0.1}$ & $0.9\pm0.1$         & 7.9 & $1.8\pm0.3$ & $ 13.0^{+0.2}_{-0.3}$ \\

NL AGN                    & 110   & $^* 44.8\pm 0.2$ & $0.8^{+0.1}_{-0.2}$ & 6.5 & $1.1^{+0.4}_{-0.3}$ & $ 12.7^{+0.3}_{-0.5}$ \\

$L_{WISE}$-selected AGN      & 127   & $^* 44.6^{+0.3}_{-0.1}$ & $0.9^{+0.1}_{-0.2}$ & 14.1 & $1.6\pm0.3$ & $ 12.9\pm 0.2$ \\

NL + $L_{WISE}$-selected AGN & 237  & $^* 44.7^{+0.3}_{-0.1}$ & $0.9^{+0.1}_{-0.2}$ & 13.5 & $1.2\pm0.3$ & $ 12.6^{+0.3}_{-0.5}$ \\

\hline \hline
\end{tabular}

\vspace{0.05in}
\raggedright
\footnotesize
      $^*$ Luminous AGN sample ($L_{6\mu m} > 3\times 10^{44}$ erg s$^{-1}$)

\label{tab:SummMass} 
\end{table*}

In this section, we discuss the measurement of the projected real-space correlation function of DESI galaxies. First we present the LRG autocorrelation, then we turn to the AGN-LRG cross-correlations.

\subsection{DESI LRG Autocorrelation}\label{sec:autocorr}

In order to estimate the bias and halo mass for our AGN populations in a cross-correlation with the DESI LRG sample, we must first calculate the autocorrelation of the LRGs. These have an average number density of 620 deg$^{-2}$. Estimating the two-point correlation function $w_p(r_p)$ following the methods outlined in \S \ref{sec:corrfunc}, we show the measured per-field clustering statistic in Figure \ref{fig:autocorr}. The fitted linear DM halo models are shown with dashed lines; they depend principally on the median $z$ of the sample and weakly on the shape of $dN/dz$. We do not fit the correlation function interior to $1 h^{-1}$ Mpc because our model does not consider the intra-halo clustering driven by the 1-halo term. These per-field correlation functions agree within the $2\sigma$ uncertainties. The two-halo term regime of the correlation functions follow theoretical and previously measured expectations out to the largest scales ($r_p > 20 h^{-1}$ Mpc) that we can measure.  

We inverse-variance weight the inferred galaxy bias from each field, and recover an average LRG linear bias of $b_G = 2.3^{+0.1}_{-0.2}$, consistent with prior spectroscopic LRG clustering analyses at this redshift \citep[c.f.][]{zhai_clustering_2017}. Fitting only the points with $r_p > 5\,h^{-1}\,$Mpc yields consistent values, but with larger uncertainties. We also test whether our results change when excluding the larger angular scales, and find that the measured bias values shift by $<1\sigma$ when fitting on scales $1<r_p<10 \,h^{-1}\,$Mpc. The properties of this correlation function and the LRG sample are summarized in the top row of Table \ref{tab:SummMass}. With the galaxy bias in hand, we now turn to the cross-correlations with AGN. 

\subsection{LRG × AGN Cross-correlations}

With the sample of AGN as defined in \S \ref{sec:classification}, we will now use the projected two-point correlation function to measure the spatial clustering of these objects. The normalized redshift distribution of these different AGN sub-samples is shown in Figure \ref{fig:dndz}.

By measuring the clustering of each AGN sample, we can infer if there are any appreciable differences in the clustering strength of the galaxies relative to our DM halo model, and infer the average halo mass for the sample following the procedure outlined in \S \ref{sec:model_def}. We first calculate the cross-correlation between the DESI LRGs and the full AGN sample in each of our three HSC$\times$DESI overlap fields. This is illustrated in the top left panel of Figure \ref{fig:crosscorr}. 

We perform the cross-correlation analysis for AGN that are in $z\in 0.5-1.0$ and lie above a luminosity threshold of $L_{6 \,\mu m} > 3\times 10^{44} \,{\rm erg \, s^{-1}}$, as established in \S \ref{sec:classification}. The measured $w_p(r_p)$ functions across the three fields are consistent with each other within their $1\sigma$ per-bin uncertainties, but note that they are considerably noisier than the autocorrelation. This is expected given the dramatic drop in number of objects in the analysis from $\sim27,000$ LRGs to 704 AGN in the clustering sample, for an average number density of 16.2 deg$^{-2}$. The open symbols in Figure \ref{fig:crosscorr} represent the inverse-variance weighted per-bin mean of the correlation functions across the three fields, while the dashed lines represent the DM halo model fit to the measured $w_p(r_p)$. We overlay the average DM halo model as a black dot-dashed line, and reproduce it as a reference for all the AGN sub-samples' correlation functions illustrated in Figure \ref{fig:crosscorr}. Fitting the linear bias for the two-halo term physical scales $r_p > 1\, h^{-1}$ Mpc, and dividing by the contribution of the LRGs we estimated in \S \ref{sec:autocorr}, we find that the average bias of the full AGN sample is $\langle b_A \rangle = 1.7 \pm 0.2$, having assumed no stochastic scatter term \citep{dekel_stochastic_1999, tegmark_observational_1999}. 

We then use the average halo mass inference formalisms adopted in \S \ref{sec:model_def}. We incorporate the \cite{tinker_large-scale_2010} halo mass function and the measured $dN/dz$ of the sample and we infer that the average halo mass for the full AGN sample is $\langle M_h \rangle = 13.1^{+0.1 }_{-0.2}\,\log(h^{-1}\,M_\odot)$. This value is consistent with prior inferences of the typical AGN halo mass \citep{cappelluti_clustering_2012, shen_cross-correlation_2013, krumpe_more_2014, timlin_clustering_2018}, and with our value in \cite{cordova_rosado_cross-correlation_2024}.

We now turn to the individual AGN subsample's measured correlation functions and inferred average halo mass. The corresponding $dN/dz$ for each of the subsamples considered in this analysis is represented in Figure \ref{fig:dndz}. These (luminosity and redshift-constrained) subsamples have average densities of 10.8 deg$^{-2}$ for the BL AGN, 2.5 deg$^{-2}$ for the NL AGN, and 2.9 deg$^{-2}$ for the $L_{WISE}$-selected AGN. We also split the BL sample into those with a steeper $\beta \leq -1$ having 7.0 deg$^{-2}$, and the  obscured $\beta > -1$ BL AGN having 3.8 deg$^{-2}$. We also calculate the clustering of the joint NL + $L_{WISE}$ AGN sample, with 5.5 deg$^{-2}$. As before, we fit each of these with a linearly biased DM halo model at $r_p > 1\,h^{-1}$ Mpc, to determine the $b_G b_A$. We then divide by $b_G$ to determine the AGN bias $b_A$. We then infer the halo mass from this $b_A$ and the sample $dN/dz$ following \S \ref{sec:model_def}. Following the same procedure as above, we calculate the projected two-point correlation function for each AGN sub-type sample, which we show in Figure \ref{fig:crosscorr}. These cross-correlations display a fair amount of field-to-field variation, but are consistent within the errors. We account for these variations and illustrate a mean value as we do for the full AGN sample.

\begin{figure*}
    \centering
    \includegraphics[width = 1.\linewidth]{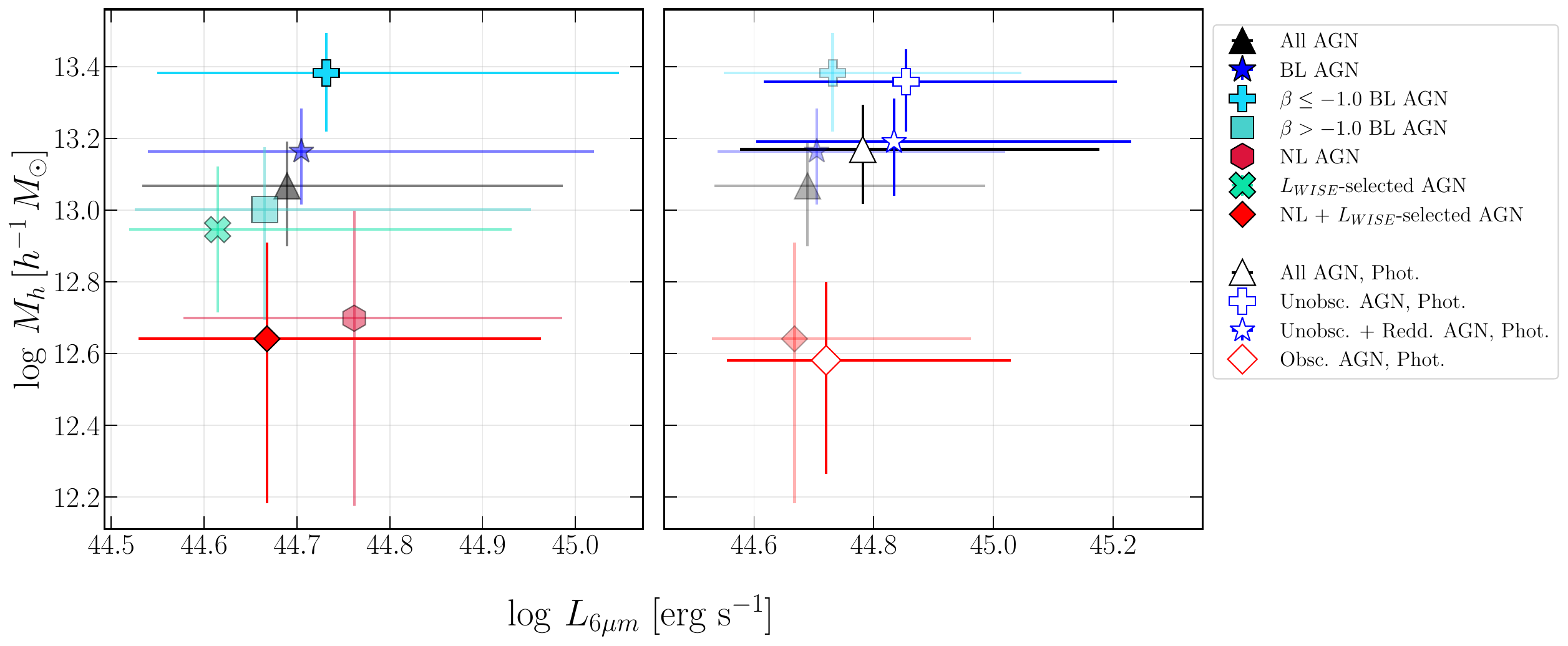}
    \caption{Summary of recovered halo masses ($M_h$) from the spectroscopic AGN cross-correlations, and in comparison with our result from the photometric cross-correlation from \cite{cordova_rosado_cross-correlation_2024}. \textit{Left:} The $\langle M_h \rangle$ for each of our spectroscopic samples, as a function of the median $L_{6\mu m}$ of the sample. We highlight the difference between the inferred $M_h$ of the blue ($\beta \leq -1$) BL and NL + $L_{WISE}$-selected AGN (the two opaque symbols), where the former are in DM halos that are $5.5\times$ more massive than the latter, at $2.8\sigma$ statistical significance. The values represented here are also found in Table \ref{tab:SummMass}, and the symbols for each (sub)sample here match those in Figure \ref{fig:crosscorr}. \textit{Right:} The $\langle M_h \rangle$ results from \cite{cordova_rosado_cross-correlation_2024} with the uncertainty from the full covariance matrix estimation (open symbols), overlaid on the analogous spectroscopic sample from this analysis (symbol shapes are shared). We note the $<1\sigma$ shifts in the average inferred halo mass between each of the analogous photometric and spectroscopic samples, suggesting that systematic biases in our redshifts that would affect the inferred $M_h$ are sub-dominant to statistical uncertainties.}
    \label{fig:masssumm}
\end{figure*}

We present the AGN sub-type halo masses in Figure \ref{fig:masssumm} in Table \ref{tab:SummMass}. The average halo mass for the BL AGN is $\langle M_h \rangle = 13.2\pm 0.1$ $\log(\, h^{-1}\,M_\odot)$. We find that the $\langle M_h\rangle$ for NL AGN is $12.7^{+0.3}_{-0.5}$ $\log(\, h^{-1}\,M_\odot)$. The $L_{WISE}$-selected AGN average $M_h$ is $12.9\pm0.2$ $\log(\, h^{-1}\,M_\odot)$.
When splitting the BL AGN sample by the continuum power-law slope, we find that the steeper and bluer ($\beta \leq -1$) BL AGN have $\langle M_h \rangle 
 = 13.4^{+0.1}_{-0.2}$ $\log(\, h^{-1}\,M_\odot)$, while the redder ($\beta > -1$) BL AGN have $\langle M_h \rangle 
 = 13.0^{+0.2}_{-0.3}$ $\log(\, h^{-1}\,M_\odot)$. 
Our NL + $L_{WISE}$ AGN sample has an average $M_h$ of $12.6^{+0.3}_{-0.5}$ $\log(\, h^{-1}\,M_\odot)$.

Overall, we find no statistically significant differences in the inferred halo masses between our different AGN subtype samples, but their amplitudes are informative. The apparent split in $\langle M_h \rangle$ goes in the same direction as in \cite{cordova_rosado_cross-correlation_2024}, with unobscured AGN residing in more massive halos than their obscured counterparts. But in this analysis the samples are considerably smaller, such that the uncertainties preclude us coming to any conclusions about real differences between Type I and Type II AGN. 

\section{Discussion}\label{sec:disc}

There are two outputs of this investigation: the spectroscopic inspection of $\sim 2000$ HSC+WISE-selected photometric AGN from Goulding et al. (in prep.) as observed by DESI, and the measured clustering of the HSC$\times$DESI AGN. These results allow us to better characterize the AGN populations we investigated in \cite{cordova_rosado_cross-correlation_2024}, and constrain the average halo mass for our samples having controlled for any redshift uncertainty or bias with this spectroscopic sample.

\subsection{Implications of Classification Results}\label{sec:disclass}

A critical result of this analysis is the comparison of the machine-learning based AGN selection and classification with the results from spectroscopy. \cite{hviding_spectroscopic_2024} obtained spectra of 139 objects from the HSC+\textit{WISE} AGN parent sample of Goulding et al. (in prep) using the MMT Observatory, identifying $\sim95\%$ of them as AGN ($81\%$ of the full spectroscopic sample were Type I AGN, and $14\%$ were Type II). These objects were randomly selected from the $i<22.5$ subset of the ML-defined AGN sample. %, while our HSC$\times$DESI AGN sample is affected by DESI selection effects that prefer brighter obscured AGN, as shown in Figure \ref{fig:zvsImag}. 
\cite{hviding_spectroscopic_2024} confirmed the classification of the high number density of AGN in the HSC+\textit{WISE} sample. This gives us confidence in the high number density of obscured or Type II AGN ($\sim 211$ deg$^{-2}$) that Goulding et al. (in prep) have identified, where \cite{hviding_spectroscopic_2024} have inferred number densities of $\sim 174$ deg$^{-2}$ Type I and  $\sim 170$ deg$^{-2}$ Type II AGN across $z\in0.1-3.0$. These numbers can be compared with the \cite{assef_wise_2018} R90 \textit{WISE} AGN catalog, which found a number density of $\sim 151$ deg$^{-2}$ for all AGN in their survey. The DESI spectroscopic sample provide a larger comparison sample, so we seek to understand our inferred number density for the HSC AGN catalog.

In our sample of 1991 HSC$\times$DESI ($z\in 0.5-1.0$) objects, $76\%$ are spectroscopically classified as AGN. We find that $93\%$ of the HSC$+$\textit{WISE} unobscured objects are spectroscopically classified as BL AGN ($3\%$ as NL AGN, $4\%$ are unclassified), while $59\%$ of the reddened objects are BL AGN ($12\%$ are NL AGN, $1\%$ are $L_{WISE}$-selected AGN, and $28\%$ are unclassified). Of the obscured AGN sample in the HSC+\textit{WISE} selection, we find that $13\%$ are BL AGN, $28\%$ are NL AGN, $18\%$ are $L_{WISE}$-selected AGN, and $41\%$ are unclassified. As shown in the bottom right panel of Figure \ref{fig:ohno}, those objects that are identified as NL, $L_{WISE}$-selected, or unclassified are preferentially dimmer (dominating the sample for $i> 21$) than the BL AGN (which dominate for $i<21$). 

These results are illustrative of the prevalence of BL, or Type I, AGN in our sample, and the preponderance of AGN in the HSC+\textit{WISE} selection, with a higher contamination rate than the $\sim 5\%$ that \cite{hviding_spectroscopic_2024} found. However, we also note \cite{hviding_spectroscopic_2024} had a magnitude limit of $i<22.5$, while the objects we have matched in this analysis extend to $i<23$. This difference in contamination could be primarily due to the low-$L$ objects for which we do not have optical evidence of AGN activity, in part because the redshift range and DESI observed wavelength range constrain our ability to use strong line diagnostics like the BPT diagram \citep{baldwin_classification_1981, kewley_theoretical_2001}. 

We infer an average number density of the parent HSC+\textit{WISE} sample of$\sim 240$ deg$^{-2}$, assuming that the fractions we've inferred for this $i<23$ sample are consistent for the deeper ($i\lesssim 25$) HSC+\textit{WISE} parent catalog. This is approximately split as $\sim 130$ deg$^{-2}$ BL AGN, $\sim 70$ deg$^{-2}$ NL AGN, and $\sim 39$ deg$^{-2}$ $L_{WISE}$-selected AGN. While \cite{hviding_spectroscopic_2024} report a number density that is $\sim 50\%$ higher -- DESI's magnitude limit undersamples the total obscured object population -- this result nonetheless suggests that the Goulding et al. (in prep) AGN catalog could provide the most complete optical/IR AGN catalog to date, identifying significant numbers of BL and NL AGN that have been missed in prior surveys.

The population of objects for which we do not have optical evidence of AGN activity requires further study. They have systematically lower luminosities and thus fainter magnitudes (see right panels of Figure \ref{fig:ohno}) than our BL AGN sample. When we stack the DESI spectra of the unclassified objects, we find significant [NeV] emission, which is more pronounced for the upper $L_{6\mu m}$ $50\%$ of objects. This suggests that a significant fraction of unclassified objects are likely AGN, and merit deeper observations to accurately characterize their properties.  Moreover, our designation of the $L_{WISE}$-selected AGN as active galaxies hinges on a series of assumptions about galaxy SED shape and SFR scaling relations. The MIR SEDs from \textit{WISE} photometry are sparse, and many have optical spectra with low S/N. Also, our current assumption of a canonical AGN MIR power-law to fit the \textit{WISE} photometry could be inconsistent with myriad galaxies that display significant structure in the MIR, such as the sharp features from vibrational excitation of polycyclic aromatic hydrocarbons. Additional spectroscopic follow-up for a subset of these $L_{WISE}$-selected AGN (and the unclassified objects from our matched sample) is required to understand their nature. Deeper spectroscopy and wider spectroscopic surveys that target the obscured AGN population are critical to understand this active galaxy population that has been hitherto undercounted in AGN catalogs, and remains unobserved in present DESI spectroscopic surveys.

\subsection{Interpretation and Comparison of Clustering Results}\label{sec:discuss_halomass}

While the sample and methods we use in this analysis are similar to that in our earlier work, they differ in key ways. As before, we stress that comparisons between different experiments can be affected by subtle differences in the bias--halo mass formalism, the chosen cosmological parameters, and the halo mass estimation method. While we have preserved most of our choices from \cite{cordova_rosado_cross-correlation_2024} in this analysis, we note the redshift range in our primary analysis is wider than in our earlier work, while the luminosity distribution is narrower (as a result of the significantly smaller sample). As in our earlier work, 
our selection may imprint different halo or stellar mass distributions on Type I vs Type II samples, although we have endeavored to match their luminosity distributions. Also, our prior work limited the analysis to those objects whose \textit{WISE} flux S/N was $>3$ for all \textit{WISE} bands, while the full Goulding et al. (in prep) AGN catalog that we matched with DESI in this analysis also includes objects that have S/N $>2$ in either the W2 or W3 band. Nevertheless, we find it useful to directly compare our inferred $M_h$ for the HSC+\textit{WISE} photometrically-classified AGN and this spectroscopic sample. 

The right panel of Figure \ref{fig:masssumm} compares the $\langle M_h \rangle$ of the analogous spectroscopic samples with the covariance matrix-estimated results from our photometric AGN subsample analysis ($\beta \leq -1$ BL with unobscured, BL with obscured + reddened, NL + $L_{WISE}$-selected AGN with obscured). Critically, the values are each consistent at the $<1\sigma$ level, with the only $>0.3\sigma$ shift being the full AGN sample. We selected the $\beta \leq -1$ BL AGN sample to be directly comparable with the unobscured AGN sample from our prior work. These results show that the median $\langle M_h \rangle$ inferred from our HSC$\times$DESI sample match the measurements from the wider photometric sample from HSC+WISE.  

We previously cautioned that unconstrained systematic errors in the redshift distribution of the photometric AGN could be sufficient to create the significant difference in inferred halo mass we observed between obscured and unobscured AGN (see \S 4.4 in \cite{cordova_rosado_cross-correlation_2024}). Now that we have spectroscopic redshifts, we have eliminated the potential for redshift bias for this brighter sample of AGN ($i<23$), relative to the magnitude limit in our prior work ($i<24$) for the obscured population. Though we notice the photo-$z$'s are biased relative to their DESI spec-$z$'s at $z\geq1$ among the obscured sample (see right panel of Figure \ref{fig:AGN_redshift_comp}), the effect did not lead us to infer a higher $M_h$ for the NL AGN. Rather, the median halo mass of the narrow-line AGN is consistent within $0.4\sigma$ with the $M_h$ for the photometric obscured sample, albeit with larger uncertainties. Though our spectroscopic result does not reproduce the statistical significance of our photometric result -- $2.3\sigma$ (NL)/$2.8\sigma$ (NL + $L_{WISE}$), versus $3.5\sigma$ in our prior analysis -- we expect that the full data release of the DESI Y1 data and its $10\times$ wider overlap with HSC will allow for higher precision in our measurement (up to a $6-7\sigma$ difference if the average $M_h$ is unchanged).

We refer to our extensive literature comparison in \S 5 in \cite{cordova_rosado_cross-correlation_2024} for the various optical, MIR, and X-ray selected AGN clustering studies that are useful to compare with our result, and repeat the most comparable studies here. AGN clustering experiments have produced a plethora of results; some have found Type I AGN to inhabit less massive halos than Type II AGN \citep{gilli_spatial_2009, donoso_angular_2014, dipompeo_characteristic_2017, krumpe_spatial_2018, petter_host_2023}, some find -- as we do at low significance -- that unobscured AGN are in more massive halos than obscured AGN \citep{cappelluti_active_2010, allevato_xmm-newton_2011, allevato_clustering_2014}, while others find no statistically significant difference \citep{hickox_clustering_2011, mendez_primus_2016}. Comparing with analyses that use the projected two-point correlation function \citep{shen_cross-correlation_2013, allevato_xmm-newton_2011, allevato_clustering_2014, krumpe_spatial_2018}, we agree with previously measured average halo mass constraints for the full AGN sample ($M_h \sim 12.5 - 13\,\log \,h^{-1} M_\odot$). We find no evidence suggesting that obscured/Type II AGN reside in more massive halos than Type I AGN at $z\in 0.5-1.0$.

The variety in AGN selection choices (partially depending on whether they are classified via optical or X-ray measurements) and redshift ranges make direct comparisons across analyses difficult. Nevertheless, our inferred halo mass difference between BL and NL AGN is consistent with the findings of optical selections in \cite{allevato_xmm-newton_2011} and \cite{allevato_clustering_2014}, and is inconsistent with \cite{krumpe_spatial_2018}, who find (X-ray column depth defined) Type II AGN to be in halos that are $8\times$ more massive than Type I AGN (though all at redshifts $\langle z\rangle \leq 0.03$). Upcoming full releases of optical spectroscopic surveys (c.f. DESI \citep{desi_collaboration_desi_2016}, the Prime Focus Spectrograph \citep{takada_extragalactic_2014, greene_prime_2022}) and large-area X-ray surveys like eRosita \citep{merloni_erosita_2012} will aid in disentangling these differences by revealing larger samples of AGN with which to perform these analyses. 

Given our statistical uncertainties, this analysis does not rule out a straightforward inclination explanation of the dichotomy between Type I and Type II AGN -- the ``unified model'' of AGN obscuration and emission line profiles \citep{antonucci_unified_1993, urry_unified_1995, almeida_nuclear_2017} -- to describe the halo mass difference we have measured. Taking our prior result together with spectroscopic confirmation of the average halo mass for the different AGN types, we interpret the (low statistical significance) halo mass difference we have found in this paper as suggestive that optical and MIR classifications of AGN trace different galaxy populations with their own range of properties not captured by a single geometric explanation, but larger spectroscopic samples are necessary to strengthen this result. Nevertheless, this scenario raises interesting questions about the role of mergers and potential evolutionary sequences that could better describe the differences we observe in spectroscopic properties and halo mass. An alternative description of the AGN spectroscopic classes may include hierarchical formation processes, blowouts of the obscuring material around the broad-line region, or other more exotic phases as suggested by the spectra of active galaxies recently discovered at very high redshift with \textit{JWST} \citep{labbe_uncover_2023, matthee_little_2024, greene_uncover_2024}. 

Upcoming work to better parameterize the properties of the 1-halo term is necessary to gauge the characteristics of intra-halo clustering among these AGN populations. With models that physically capture both the linear and non-linear clustering terms, we may have the tools to address further questions on the distribution of AGN in DM halos and clusters, and provide observational constraints for simulations that connect galaxy formation and large scale structure. Small-scale models are particularly necessary in the context of understanding AGN feedback in cluster environments and its effect on large scale structure and cosmological parameter inference \citep{fabian_observational_2012, haggard_field_2010, amon_non-linear_2022, koulouridis_agns_2024}. Upcoming surveys will extend the redshift range of this sort of analysis, providing the means to investigate whether the inferred halo mass difference persists across cosmic epochs, with its own implications on AGN evolution and formation histories.

\section{Conclusion}\label{sec:conclu}

We present a cross-correlation analysis between luminous red galaxies observed by DESI and active galactic nuclei matched between our parent HSC+\textit{WISE} photometry-selected sample and DESI Early Data Release observations. We fit the projected two-point correlation function at physical scales $1 < r_p < 40 \, h^{-1}\, {\rm Mpc}$. From the initial HSC optical and \textit{WISE} MIR photometric AGN selection, we find $\sim 4,222$ matched objects in the GAMA field of HSC in the DESI EDR, and we visually inspect and classify the $1991$ objects in the $z \in 0.5-1.0$ range. With a total of $43.4$ deg$^{2}$ of overlap, we have an AGN number density of 16.2 deg$^{-2}$ after limiting to all objects with $L_{6\mu m} > 3\times 10^{44}$ erg s$^{-1}$. Cross-correlating with $\sim 27,000$ DESI LRGs, we measure the $w_p (r_p)$ clustering statistic for 704 spectroscopically-confirmed and classified AGN, splitting it into AGN subtypes. We fit these correlation functions with a linear DM halo model at physical scales $r_p > 1\, h^{-1}{\rm Mpc}$, and interpret the clustering strength with physical parameters. Our principal conclusions are as follows.

\begin{enumerate}
    \item We visually inspect and classify 1,991 HSC$\times$DESI objects. We confirm that $76\%$ are galaxies hosting an AGN. Of these 1,517 quasars, $74\%$  are BL AGN, $19\%$ are NL AGN, and $9\%$ are the $L_{WISE}$-selected AGN. Based on these classification fractions, we infer that the scaled number density of the complete HSC+\textit{WISE} AGN catalog ($z\in 0.1-3$) is at least $\sim 240$ deg$^{-2}$. This is split into $\sim 130$ deg$^{-2}$ BL AGN, $\sim 70$ deg$^{-2}$ NL AGN, and $\sim 39$ deg$^{-2}$ $L_{WISE}$-selected AGN. The high rate of AGN confirmation and the estimated number densities suggest that our parent HSC+\textit{WISE} AGN sample is one of the most complete optical/IR AGN catalog to date. 

    \item The host halo masses of unobscured ($\beta \leq -1$) BL AGN ($\langle M_h \rangle = 13.4^{+0.1}_{-0.2}$ $\log(\, h^{-1}\,M_\odot)$) are found to be $\sim 5.5\times$ more massive than the halos that host NL + $L_{WISE}$-selected AGN ($\langle M_h \rangle = 12.6^{+0.3}_{-0.5}$ $\log(\, h^{-1}\,M_\odot)$), at $2.8\sigma$ statistical significance. The central values are entirely consistent with their analogous photometric AGN classification we previously measured at higher precision. While this result is not independent evidence that this difference in characteristic halo mass is real, it does suggest that our earlier findings of a statistically significant difference are not strongly impacted by systematic biases.

\end{enumerate}

We have studied the relationship between galaxies and the accreting SMBHs they host. We find that the averaged inferred DM halo masses that host these active galaxies continue to be a rich statistic with which to trace AGN properties and evolutionary histories. In performing a projected two-point correlation function of a luminosity-constrained sample, we recover the clustering strength differences between AGN sub-types. We do not, however, attempt to fit nor infer the properties of the intra-halo clustering ($r_p < 1\, h^{-1}$ Mpc) at this stage of our investigation. Additional work is needed to exploit these small scales with an HOD-style treatment of the intra-halo clustering. Independent constraints of the halo mass via gravitational lensing experiments with CMB lensing or galaxy-galaxy lensing by the different AGN samples can also serve as an independent test of the halo masses we have constrained here. Development of these analytical tools is paramount to understand AGN clustering properties in this new era of wide-field cosmological galaxy surveys by experiments like the Rubin Observatory's Legacy Survey of Space and Time \citep{ivezic_lsst_2019}, Euclid \citep{amendola_cosmology_2018}, the Prime Focus Spectrograph \citep{takada_extragalactic_2014, greene_prime_2022} and the Dark Energy Spectroscopic Instrument \citep{desi_collaboration_desi_2016, desi_collaboration_desi_2024}. In creating these analytic frameworks and working to build robust AGN catalogs, we will have the tools to better understand the evolutionary history of AGN across cosmic epochs.

\section*{Acknowledgments}

We thank P. Melchior, C. Ward, L. A. Perez, and J. Givans for helpful conversations throughout the course of this work. 

Computing was performed using the Princeton Research Computing resources at Princeton University. R.C.R. acknowledges support from the Ford Foundation Predoctoral Fellowship from the National Academy of Sciences, Engineering, and Medicine. ADG and RCR gratefully acknowledge support from the NASA Astrophysics Data Analysis Program \#80NSSC23K0485. ADG and JEG acknowledge support from the National Science Foundation under Grant Number AST-1613744, and JEG acknowledges support from the National Science Foundation under Grant Number AST-2306950.

The Hyper Suprime-Cam (HSC) Collaboration includes the astronomical communities of Japan and Taiwan, and Princeton University. The HSC instrumentation and software were developed by the National Astronomical Observatory of Japan (NAOJ), the Kavli Institute for the Physics and Mathematics of the Universe (Kavli IPMU), the University of Tokyo, the High Energy Accelerator Research Organization (KEK), the Academia Sinica Institute for Astronomy and Astrophysics in Taiwan (ASIAA), and Princeton University. Funding was contributed by the FIRST program from the Japanese Cabinet Office, the Ministry of Education, Culture, Sports, Science and Technology (MEXT), the Japan Society for the Promotion of Science (JSPS), Japan Science and Technology Agency (JST), the Toray Science Foundation, NAOJ, Kavli IPMU, KEK, ASIAA, and Princeton University. 

This paper makes use of software developed for Vera C. Rubin Observatory. We thank the Rubin Observatory for making their code available as free software at \url{http://pipelines.lsst.io/}. This paper is based on data collected at the Subaru Telescope and retrieved from the HSC data archive system, which is operated by the Subaru Telescope and Astronomy Data Center (ADC) at NAOJ. Data analysis was in part carried out with the cooperation of Center for Computational Astrophysics (CfCA), NAOJ. 

We are honored and grateful for the opportunity of observing the Universe from Maunakea, which has the cultural, historical and natural significance in Hawaii. 

This material is based upon work supported by the U.S. Department of Energy (DOE), Office of Science, Office of High-Energy Physics, under Contract No. DE–AC02–05CH11231, and by the National Energy Research Scientific Computing Center, a DOE Office of Science User Facility under the same contract. Additional support for DESI was provided by the U.S. National Science Foundation (NSF), Division of Astronomical Sciences under Contract No. AST-0950945 to the NSF’s National Optical-Infrared Astronomy Research Laboratory; the Science and Technologies Facilities Council of the United Kingdom; the Gordon and Betty Moore Foundation; the Heising-Simons Foundation; the French Alternative Energies and Atomic Energy Commission (CEA); the National Council of Science and Technology of Mexico (CONACYT); the Ministry of Science and Innovation of Spain (MICINN), and by the DESI Member Institutions: \url{https://www.desi.lbl.gov/collaborating-institutions}. Any opinions, findings, and conclusions or recommendations expressed in this material are those of the author(s) and do not necessarily reflect the views of the U.S. National Science Foundation, the U.S. Department of Energy, or any of the listed funding agencies. 

The authors are honored to be permitted to conduct scientific research on Iolkam Du’ag (Kitt Peak), a mountain with particular significance to the Tohono O’odham Nation.

\facilities{Subaru (HSC), WISE, NEOWISE, Mayall (DESI), Sloan}

\software{\texttt{Astropy} \citep{astropy_collaboration_astropy_2018, astropy_collaboration_astropy_2022}, \texttt{Matplotlib} \citep{hunter_matplotlib_2007}, \texttt{NumPy} \citep{van_der_walt_numpy_2011, harris_array_2020}, \texttt{SciPy} \citep{virtanen_scipy_2020}}, \texttt{Corrfunc} \citep{sinha_corrfunc_2020}, Core Cosmology Library \citep{chisari_core_2019}

\bibliographystyle{aasjournal}
\bibliography{references, references2}

\end{document}